\documentclass[a4paper,11pt]{article}
\pdfoutput=1
\usepackage{jcappub}
\usepackage{mathrsfs,amsmath,latexsym,amssymb,amsfonts,epsfig,cancel,enumerate,graphicx,txfonts,subfigure,graphics,diagbox}
\usepackage{multirow}
\usepackage{booktabs}
\usepackage{xcolor}

\usepackage{etoolbox}
\usepackage{orcidlink}
\title{\boldmath Electric Penrose process in the spacetime of a quantum-corrected Reissner-Nordstr\"om black hole}

\author[a]{Jiawei Chen\orcidlink{0009-0003-5390-8186},}
\author[a,1]{Jinsong Yang\orcidlink{0000-0003-4051-2767}\note{Corresponding author.}}

\affiliation[a]{School of Physics, Guizhou University, Guiyang 550025, China}

\emailAdd{gs.chenjw23@gzu.edu.cn}
\emailAdd{jsyang@gzu.edu.cn}

\abstract{In this paper, we study the electric Penrose process for charged particles in the spacetime of a covariant quantum-corrected Reissner-Nordstr\"om black hole. We first derive the equations of motion for charged particles around the black hole, and then analyze how the quantum parameter $\zeta$ modifies the generalized ergoregion boundary and affects the energy-extraction efficiency. We further analyze the subsequent motion of charged particles in the electric Penrose process, and rigorously prove that under specific simplified conditions, the resulting fragment particle can always escape to a distant observer with a net energy gain, a conclusion applicable to a wide range of charged black hole models. Finally, we study the electric Penrose process in a critical regime where the initial particle is bound, yet its high-energy fragment particle may still escape. A key finding is that while $\zeta$ slightly alters the particle trajectories, it can qualitatively alter outcomes near critical conditions, causing a fragment particle that would escape in the classical black hole spacetime to become trapped in the quantum-corrected one. These results collectively demonstrate the obstructive effect of quantum corrections on the Penrose process and provide potential kinematic signatures to distinguish the quantum-corrected from classical Reissner-Nordstr\"om black holes.}

\keywords{absorption and radiation processes, modified gravity, quantum black holes}

\arxivnumber{2601.01508}

\begin{document}

\maketitle
\flushbottom

\section{Introduction}

Black holes (BHs), predicted by general relativity, have now been confirmed through various observational means~\cite{LIGOScientific:2016aoc,EventHorizonTelescope:2022wkp}. Their intense gravitational fields make them ideal laboratories for testing diverse gravitational theories. Simultaneously, BHs possess enormous energy. Phenomena such as high-energy radiation and relativistic jets produced by the accretion disks surrounding BHs~\cite{Falcke:1993kd} constitute some of the most active and energetically efficient astrophysical processes in the Universe. It was generally believed that this immense energy was inaccessible to external extraction. This view changed when Penrose, based on the existence of negative-energy orbits within the ergoregion, proposed a mechanism for extracting energy from a rotating BH, later known as the Penrose process~\cite{Penrose:1971uk}. This proposal fundamentally transformed our understanding of BHs, redefining them from passive gravitational endpoints into active media capable of exchanging energy with their surroundings.

The Penrose process~\cite{Penrose:1971uk} can be outlined as follows: a test particle 1 with energy $E_1$ ($>0$) falls freely from infinity into the ergoregion of a rotating BH. At a certain spacetime point, it splits into two fragments-particle 2 and particle 3. By choosing appropriate splitting conditions, particle 2 can be placed on a negative-energy orbit and eventually falls into the BH. According to the equivalence principle of general relativity, 4-momentum is conserved in curved spacetime, i.e., $P^{a}_1 = P^{a}_2 + P^{a}_3$. Consequently, particle 3 carries a higher energy $E_3$ ($> E_1$) and returns to infinity along an outgoing geodesic. The energy extracted from the BH in this process is $E_3 - E_1$; theoretical analyses show that up to about $29\%$ of the total energy of the rotating BH can be extracted~\cite{Wagh:1989zqa}. To date, the Penrose process and its various extensions remain an active research field, continually attracting broad theoretical and observational interest~\cite{Patel:2023efv,Zaslavskii:2023jxu,Ruffini:2024dwq,Viththani:2024map,Kar:2025anc,Zhao:2025ouq,Wang:2025nfv}. It should be noted that, in general, the above Penrose process relies on the rotation of the BH. Since static neutral BHs lack ergoregion, they do not support negative-energy orbits, rendering energy extraction via this mechanism generally impossible.

However, subsequent theoretical studies revealed that for static charged BHs, such as the Reissner-Nordstr\"om (RN) BH, a charged particle with a sign opposite to that of the BH's charge can occupy a negative-energy orbit in the vicinity of the BH~\cite{Wagh:1989zqa,Denardo:1973pyo,Denardo:1974qis,Dadhich:1978gg}. This crucial finding implies that energy extraction from a static charged BH is possible via electromagnetic interaction between the charges, without relying on the BH's rotation. This mechanism, known as the electric Penrose process, extends the framework of energy extraction from rotating BHs to static charged ones. Subsequently, research has not only examined the general properties of the electric Penrose process~\cite{Kokubu:2021cwj,Zaslavskii:2024zgh} but has also extended to investigate charged binary BHs~\cite{Sanches:2021kye,Baez:2022dqu}, charged BHs in various modified-gravity theories~\cite{Tursunov:2021jjf,Vertogradov:2022eeq,Xamidov:2024xpc,Alloqulov:2024cto,Vertogradov:2025wgg}, and charged BHs with a cosmological constant~\cite{Feiteira:2024awb,Baez:2024lhn,Feiteira:2024dwn,Chen:2025bph}. These studies collectively reveal the influence of charge, magnetic fields, and the cosmological constant on energy extraction. Furthermore, recent studies have demonstrated that a BH's charge cannot be completely discharged to zero through repeated application of this classical process, with significant energy dissipation occurring during extraction~\cite{Hu:2025bbc}. Nonetheless, the electric Penrose process remains an ideal and effective method for energy extraction, warranting further investigation.

Beyond reviewing the aforementioned extensions of the electric Penrose process, a natural and important theoretical extension is to examine its behavior within the framework of a recently proposed effective quantum-gravity theory. This theory resolves the long-standing issue of general covariance and provides covariant conditions in spherically symmetric models~\cite{Zhang:2024khj,Zhang:2024ney}, attracting significant attention~\cite{Konoplya:2024lch,Liu:2024soc,Liu:2024wal,Li:2024afr,Wang:2024iwt,Skvortsova:2024msa,Ban:2024qsa,Du:2024ujg,Lin:2024beb,Shu:2024tut,Liu:2024pui,Cafaro:2024lre,Konoplya:2025hgp,Chen:2025ifv,Ai:2025myf,Lutfuoglu:2025hwh,Chen:2025aqh,Motaharfar:2025ihv,Sahlmann:2025fde,Zhang:2025ccx,Calza:2025mwn,Umarov:2025wzm,Liu:2025iby,Huang:2025gia,Du:2025kcx,Liu:2025hcx,Chen:2026kbn}. Subsequently, the framework was further extended to the electrovacuum case with a cosmological constant, yielding several charged quantum-corrected BH solutions~\cite{Yang:2025ufs}. This work focuses on one of these solutions and considers the case of $\Lambda = 0$. Our aim is to systematically investigate the influence of the quantum-gravity effects inherent in this spacetime on the electric Penrose process. We will particularly focus on the dynamics of particle motion during the process. This aspect has been insufficiently explored in the existing literature. Specifically, we will analyze how the quantum parameter $\zeta$ modulates the critical conditions and efficiency of energy extraction, and further explore the motion behavior of test particles in this process.

The paper is organized as follows. In section~\ref{section2}, we first provide a brief review on the quantum-corrected RN BH and analyze its horizon structure. We then investigate the motion of charged particles around this BH. In section~\ref{section3}, we discuss the negative-energy states of charged particles and systematically study the electric Penrose process in this quantum-corrected spacetime. In section~\ref{section4}, we examine the motion of charged particles during the general electric Penrose process. The motion of particles in a special class of the electric Penrose process is discussed in section~\ref{section5}. Finally, a summary is provided in section~\ref{section6}. Throughout this paper, we adopt geometric units with $G=c=1$, and for numerical calculations we set the BH mass $M=1$.

\section{Motion of charged particles in the spacetime of a quantum-crrected RN BH}\label{section2}

\subsection{Quantum-corrected RN BH}

Recently, a covariant quantum-corrected RN BH solution has been obtained by solving the equations of motion derived from the effective Hamiltonian constraint~\cite{Yang:2025ufs}. In the present work, we restrict our analysis to the case of vanishing cosmological constant, i.e., $\Lambda = 0$. Unless otherwise specified in the following text, the term ``quantum-corrected BH" refers specifically to the covariant quantum-corrected RN BH introduced above. In Schwarzschild coordinates, its line element is given as follows~\cite{Yang:2025ufs}:
\begin{equation}
	{\rm d}s^2=- f(r){\rm d}t^2+\frac{1}{f(r)}{\rm d}r^2+ r^2 {\rm d}\theta^2+ r^2 \sin^2{\theta} {\rm d}\phi^2,\label{metric}
\end{equation}
with
\begin{equation}
	f(r)= \left(1-\frac{2 M}{r}+\frac{Q^2}{r^2}\right)\left[1+\frac{\zeta ^2}{r^2}\left(1-\frac{2 M}{r}+\frac{Q^2}{r^2}\right)\right]. \label{metric1}
\end{equation}
The corresponding electromagnetic 4-potential $A_a$ and electromagnetic field tensor $F_{ab}$ have the following form
\begin{eqnarray}
	A_a &=&-\frac{Q}{r}({\rm d}t)_a, \label{Aa} \\
	F_{ab}&=&-\frac{Q}{r^2} ({\rm d}t)_a \wedge ({\rm d}r)_b.
	\label{AF}
\end{eqnarray}
Here, $M$, $Q$, and $\zeta$ stand for the BH mass, charge, and quantum parameter, respectively. Clearly, when $\zeta = 0$, the solution reduces to the classical RN BH solution.

The BH horizon is defined by the roots of $g^{rr} = 0$, which in this quantum-corrected BH spacetime is expressed as
\begin{equation}
	\left(1-\frac{2 M}{r}+\frac{Q^2}{r^2}\right)\left[1+\frac{\zeta ^2}{r^2}\left(1-\frac{2 M}{r}+\frac{Q^2}{r^2}\right)\right]=0. \label{horizon}
\end{equation}
We find that when $Q < M$, eq.~\eqref{horizon} admits at least two roots, denoting the inner ($r_-$) and outer horizons ($r_+$) as in the RN BH, namely
\begin{equation}
	r_{\pm}=M \pm \sqrt{M^2-Q^2}.
\end{equation}
Moreover, the term containing the quantum parameter in eq.~\eqref{horizon} may induce additional horizons in this quantum-corrected BH. This yields the following equation from eq.~\eqref{horizon}:
\begin{equation}
	F(r) \equiv 1+\frac{\zeta ^2}{r^2}\left(1-\frac{2 M}{r}+\frac{Q^2}{r^2}\right)=0.\label{Fr}
\end{equation}
It is straightforward to show that the root obtained from eq.~\eqref{Fr} is less than $r_+$.
\begin{figure}[h]
	\centering
	\includegraphics[width=3.2in, height=5.5in, keepaspectratio]{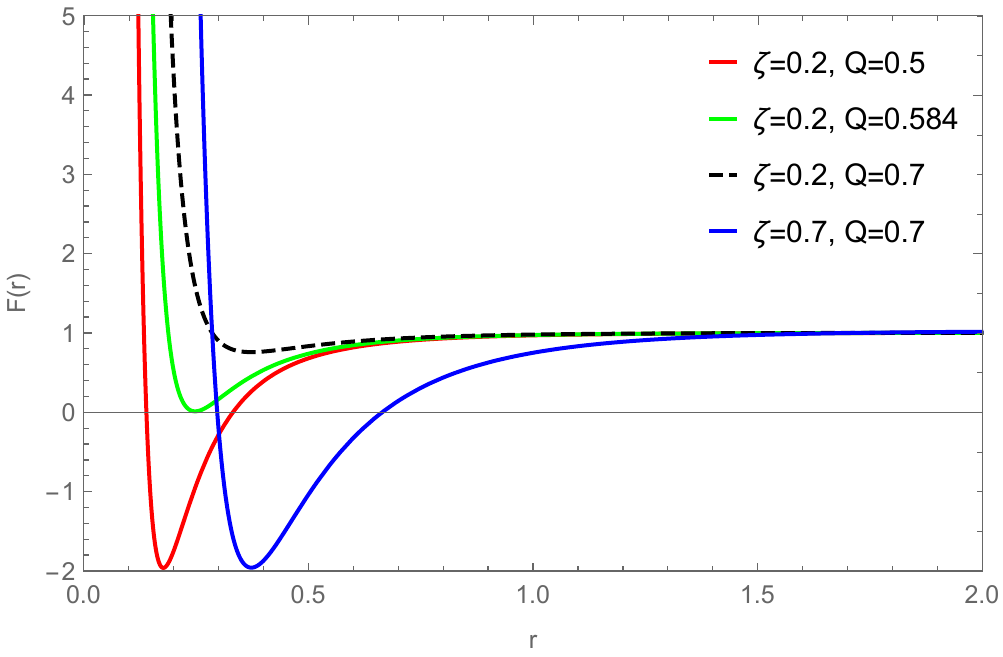}

	\caption{Behavior of the function $F(r)$ versus $r$ for different values of the parameters $Q$ and $\zeta$.}
	\label{fig_rh0}
\end{figure}

\begin{figure}[h]
	\centering
	\includegraphics[width=3.2in, height=5in, keepaspectratio]{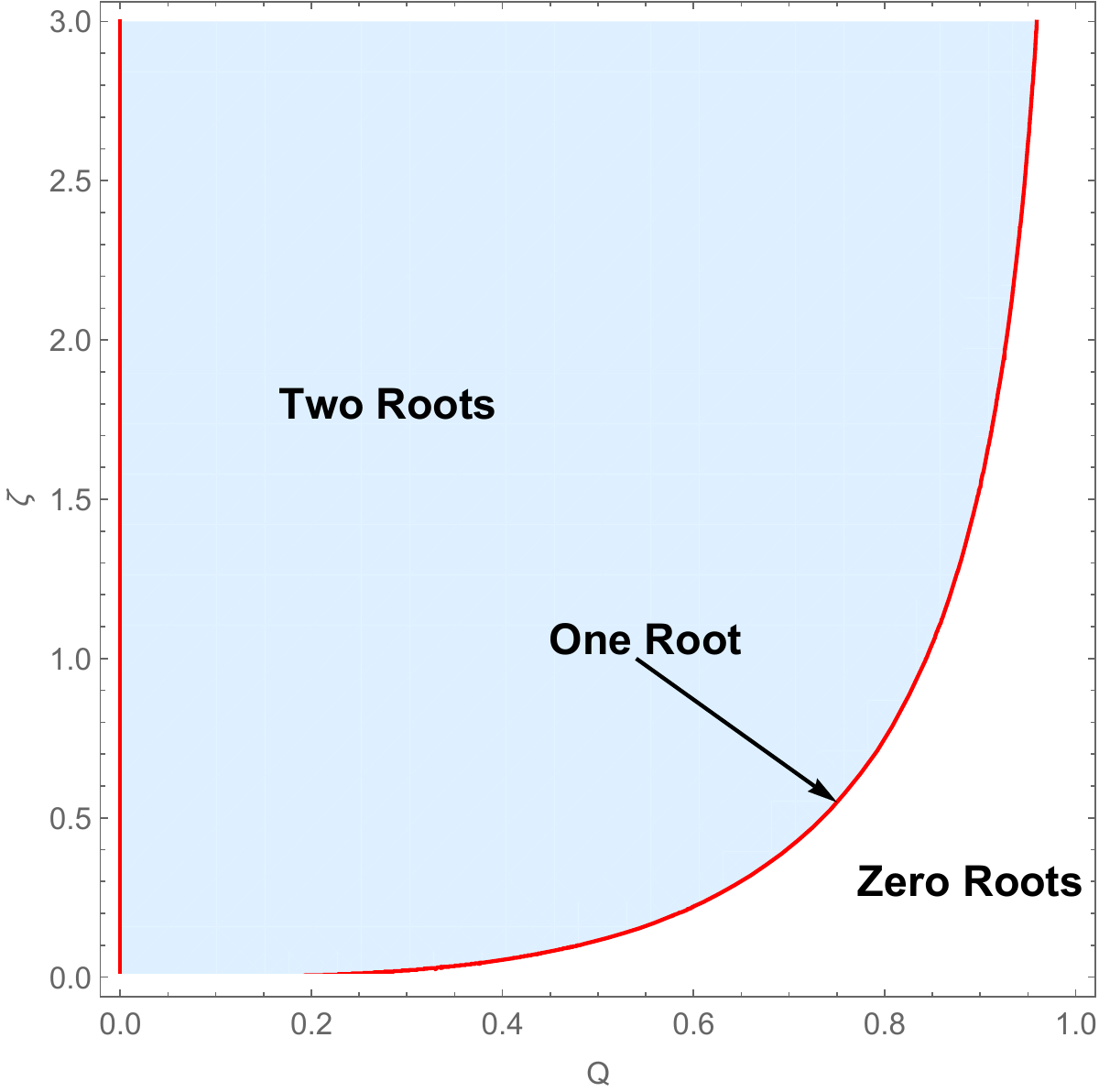}

	\caption{Root structure of $F(r)$ across parameter space $(Q,\zeta)$: one root (solid red line); two roots (blue region); zero roots (white region).}
	\label{fig_rh1}
\end{figure}

Figure~\ref{fig_rh0} delineates the behavior of the function $F(r)$ as it varies with $r$ for different parameter values, showing that $F(r)$ may exhibit zero, one, or two roots depending on the values of $\zeta$ and $Q$. Furthermore, figure~\ref{fig_rh1} illustrates the regions in the $(Q,\zeta)$ parameter space corresponding to different numbers of roots. In this figure, the two solid red lines represent the case where $F(r)$ has a single root; the blue region corresponds to two roots; and the white area indicates the case of no roots. It is noteworthy that the line $\zeta = 0$ always corresponds to the case of no roots. Therefore, this quantum-corrected BH can theoretically possess up to four horizons. Since this BH possesses no horizon larger than the outer event horizon $r_+$, our subsequent discussion will be confined to the region $r > r_+$.

\subsection{Parameter constraints and setup}

To provide a basis for the subsequent parameter selection, we recall the existing observational constraints on this quantum-corrected BH. The parameter range of this quantum-corrected BH under the M87* observational constraints is given by~\cite{Li:2026ogy}
\begin{align}
	0 \le \zeta \lesssim 2.304 \,,~~
	0 \le Q \lesssim 0.681 \,,
		\label{M87_yueshu}
\end{align}
and the corresponding one for Sgr A* is~\cite{Li:2026ogy}
\begin{align}
	0 \le \zeta \lesssim 2.875 \,, ~~ 0 \le Q \lesssim 0.799 \,.
	\label{Sgr_yeushu}
\end{align}
The above constraints naturally raise the question: can a realistic astrophysical BH truly have such a charge and quantum parameter?

In fact, various theoretical models suggest that the net charge of an astrophysical BH is extremely small~\cite{Padilla:2023bgy}, far below the theoretical upper limit and close to neutrality, due to rapid neutralization by the abundant plasma surrounding the BH~\cite{Zajacek:2018ycb}. Moreover, current gravitational-wave and Event Horizon Telescope (EHT) observations have not found significant evidence of BH charge~\cite{EventHorizonTelescope:2021dqv,Gu:2023eaa}. Therefore, although the upper bounds on $Q$ (e.g., $Q \lesssim 0.681$ and $Q \lesssim 0.799$) are theoretically allowed, they are likely to be further suppressed to much smaller values, or even zero, in realistic situations.

However, this does not render the relevant discussions meaningless. Even a tiny net charge of the BH can significantly influence its surrounding electromagnetic processes (e.g., the motion of charged particles)~\cite{Zajacek:2018ycb,Zajacek:2019kla}. On the other hand, the quantum parameter $\zeta$ in this quantum-corrected BH is proportional to the Planck length $\sqrt{\hbar}$ and, under realistic conditions, satisfies $GM \gg \zeta$, meaning that it introduces only small corrections. Nevertheless, its influence on particle motion still deserves investigation, because it may accumulate under specific conditions or lead to qualitative changes, and thus could potentially be detected in future high-precision observations. The precise values of the charge $Q$ and the quantum parameter $\zeta$ for this quantum-corrected BH remain theoretically uncertain and are difficult to directly compare with current observational constraints.

Thus, to maintain consistency with observational results and to achieve maximally manifest effects, rather than directly exploring the possible extremely small charge and the quantum parameter, this paper adopts the largest possible values of both the charge $Q$ and the quantum parameter $\zeta$ allowed within the EHT constraints, based on the upper limits derived from current observational data. Specifically, we adopt $\zeta = 0, 1, 2$ and $Q = 0.5$ as typical parameters to systematically investigate the motion of charged particles around this quantum-corrected BH, the energy extraction efficiency of the electric Penrose process, the variation of the generalized ergoregion boundary, and the escape behavior of high-energy fragment particles.

\subsection{Charged particle motion}

We next investigate the motion of charged particles around this quantum-corrected BH. We consider a particle with mass $m$ and charge $e$ moving on the equatorial plane of the quantum-corrected BH, with the 4-velocity of the test particle given by $u^a=\frac{d x^a}{d \tau}=\dot{x}^a$. Thus, its motion is governed by the Lagrangian denstity~\cite{Wagh:1989zqa,Pugliese:2011py}
\begin{equation}
	\begin{split}
		\mathcal{L}&=\frac{1}{2} g_{a b}\dot{x}^{a}\dot{x}^{b}+ q A_a \dot{x}^{a}=\frac{1}{2}\left[- f(r)\dot{t}^{2}+\frac{1}{f(r)}\dot{r}^{2}+ r^2 \dot{\phi}^{2}\right]-\frac{qQ}{r} \dot{t}.
		\label{Lagrangian_density}
	\end{split}
\end{equation}
Here, we have utilized eqs.~\eqref{metric} and \eqref{Aa} and simplified them by considering the test particle motion on the equatorial plane with $\theta = \pi/2$. The dot denotes derivative with respect to proper time $\tau$, and $q \equiv e/m$ represents the specific charge of the charged particle.

Based on the Lagrangian density in eq.~\eqref{Lagrangian_density}, which exhibits no explicit dependence on $t$ or $\phi$, the corresponding components of the generalized momentum $P_a$ are conserved~\cite{Wagh:1989zqa}. Consequently, the conserved specific energy $E$ and specific angular momentum $L$ of the particle can be expressed as~\cite{Wagh:1989zqa,Pugliese:2011py}:
\begin{align}
	&P_t = \frac{\partial \mathcal{L}}{\partial \dot{t}}= -f(r) \dot{t} - \frac{qQ}{r}=-\frac{\bar{E}}{m} \equiv -E, \label{E}\\
	&P_\phi = \frac{\partial \mathcal{L}}{\partial \dot{\phi}}= r^2 \dot{\phi}=\frac{\bar{L}}{m}\equiv L \label{L}.
\end{align}
Here, $\bar{E}$ and $\bar{L}$ represent the energy and angular momentum of the test particle, respectively. From eqs.~\eqref{E} and \eqref{L}, we obtain
\begin{align}
	\dot{t}=\frac{E}{f(r)}-\frac{qQ}{f(r)r}, \qquad \dot{\phi}=\frac{L}{r^2}.\label{tfai}
\end{align}

Moreover, for the massive particle, its 4-velocity $u^a$ satisifies $g_{ab} u^a u^b=-1$. Thus, we have:
\begin{equation}
	\begin{split}
		-1&=g_{a b}\dot{x}^{a}\dot{x}^{b}=-f(r)\dot{t}^{2}+\frac{1}{f(r)}\dot{r}^{2}+ r^2 \dot{\phi}^{2}.\label{timelike}
	\end{split}
\end{equation}
Inserting eq.~\eqref{tfai} into eq.~\eqref{timelike}, we derive
\begin{equation}
	\begin{split}
		\dot{r}^2&=\left(E-\frac{qQ}{r}\right)^2-f(r)\left(1+\frac{L^2}{r^2}\right)\equiv\left(E-V_{+}\right)\left(E-V_{-}\right),\\
	\end{split}
\end{equation}
where
\begin{equation}
	\begin{split}
		V_{\pm}\equiv\frac{qQ}{r} \pm \sqrt{f(r)\left(1+\frac{L^2}{r^2}\right)}.\label{Eq:Vz}
	\end{split}
\end{equation}
The effective potential $V_{\pm}$ denotes the energy of the radial motion when $\dot{r}$ vanishes. In this paper, we focus solely on the case of $V_{+}$.

\begin{figure*}[htbp]
	\centering
	\includegraphics[width=3.2in, height=5in, keepaspectratio]{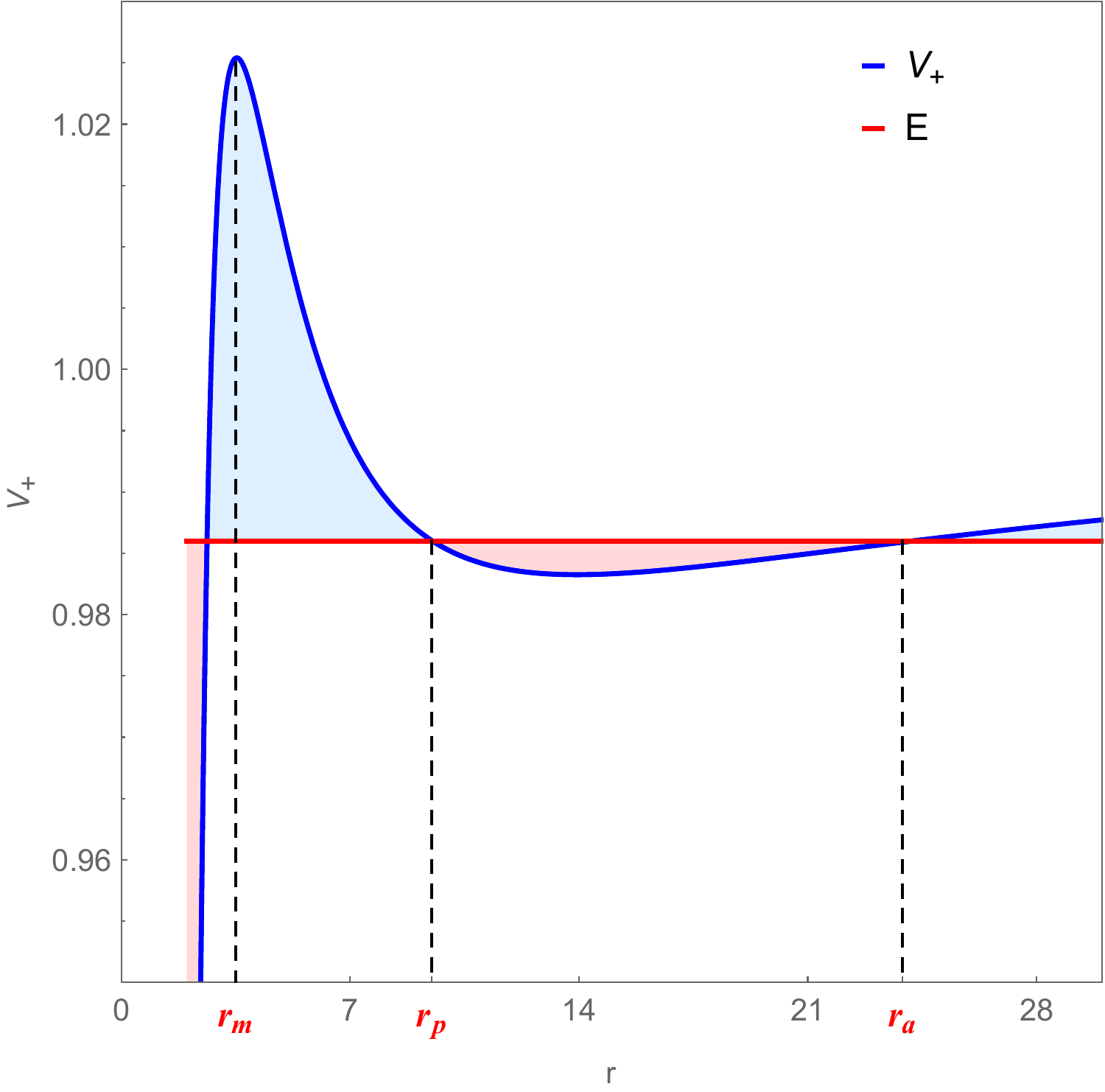}

	\caption{The relative magnitude between the effective potential curve and the particle's energy. Here, the blue and red solid curves represent $V_+$ and $E$, respectively, with the two intersection points $r_p$ and $r_a$ being the turning points. The peak of the effective potential $V_+$ occurs at the radial coordinate $r_m$.}
	\label{fig_Vz}
\end{figure*}
\begin{figure*}[htbp]
	\centering
	\subfigure{\includegraphics[width=0.33\textwidth,height=5.5cm, keepaspectratio]{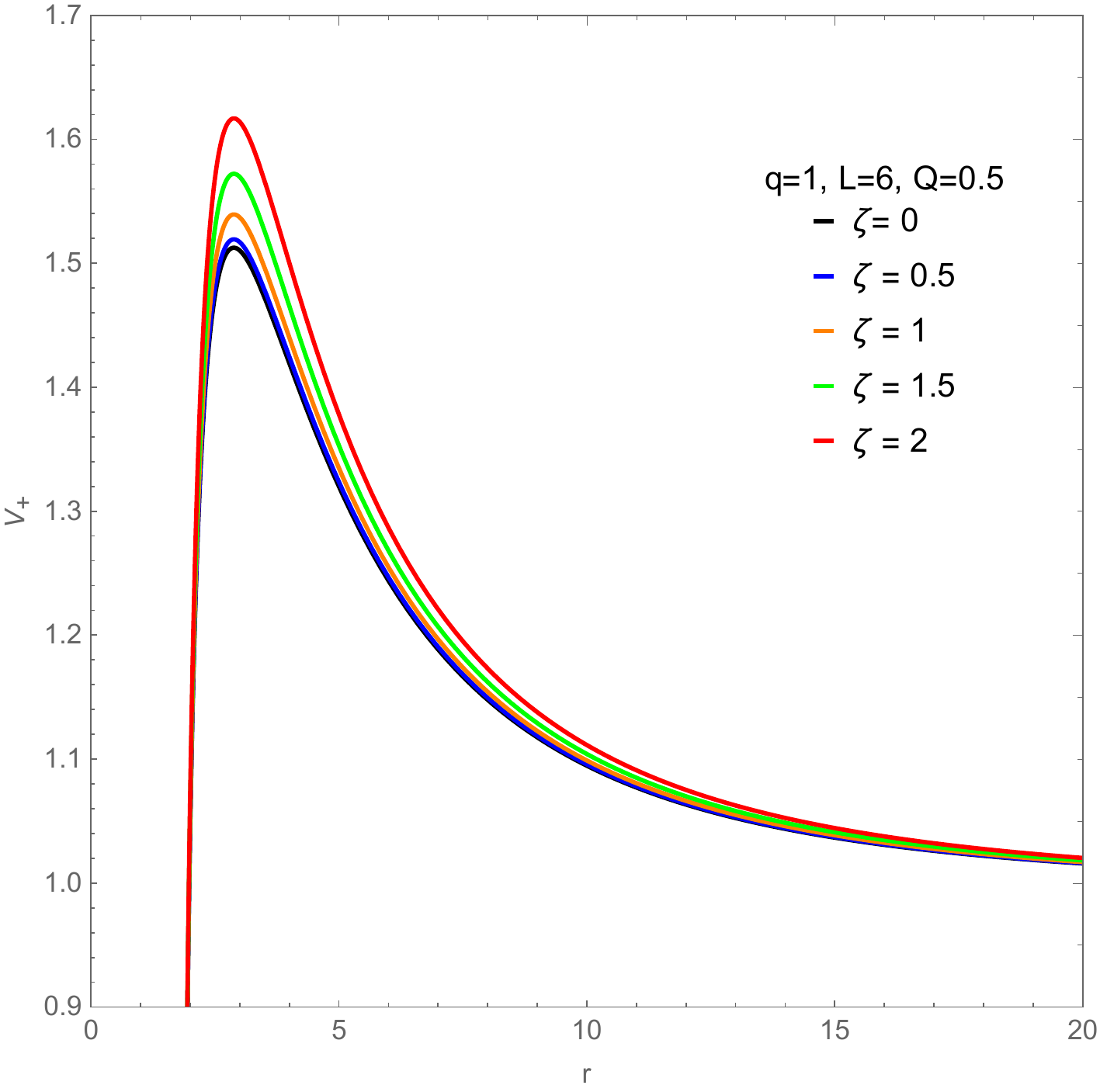}
		\includegraphics[width=0.33\textwidth,height=5.5cm, keepaspectratio]{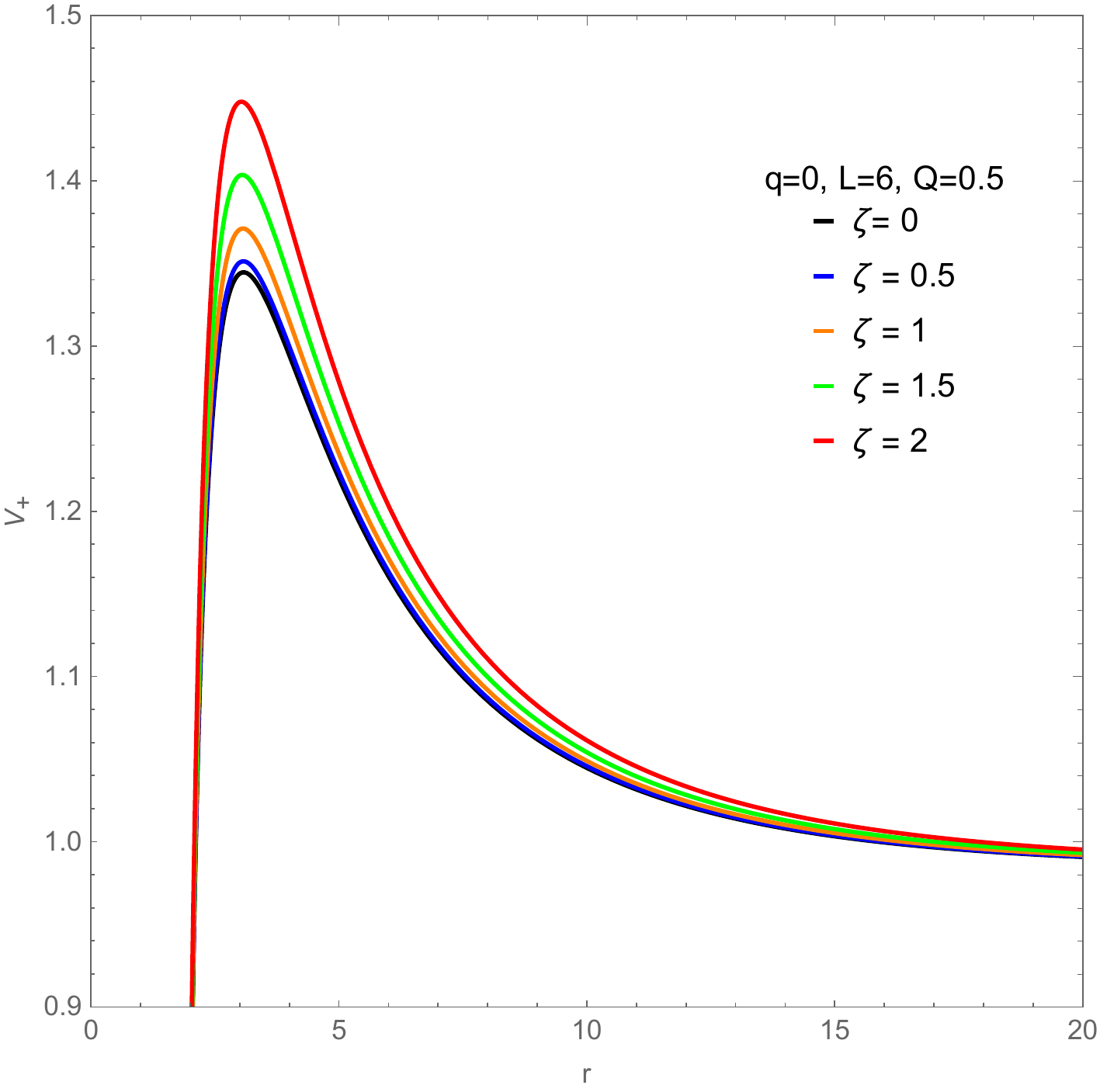}
		\includegraphics[width=0.33\textwidth,height=5.5cm, keepaspectratio]{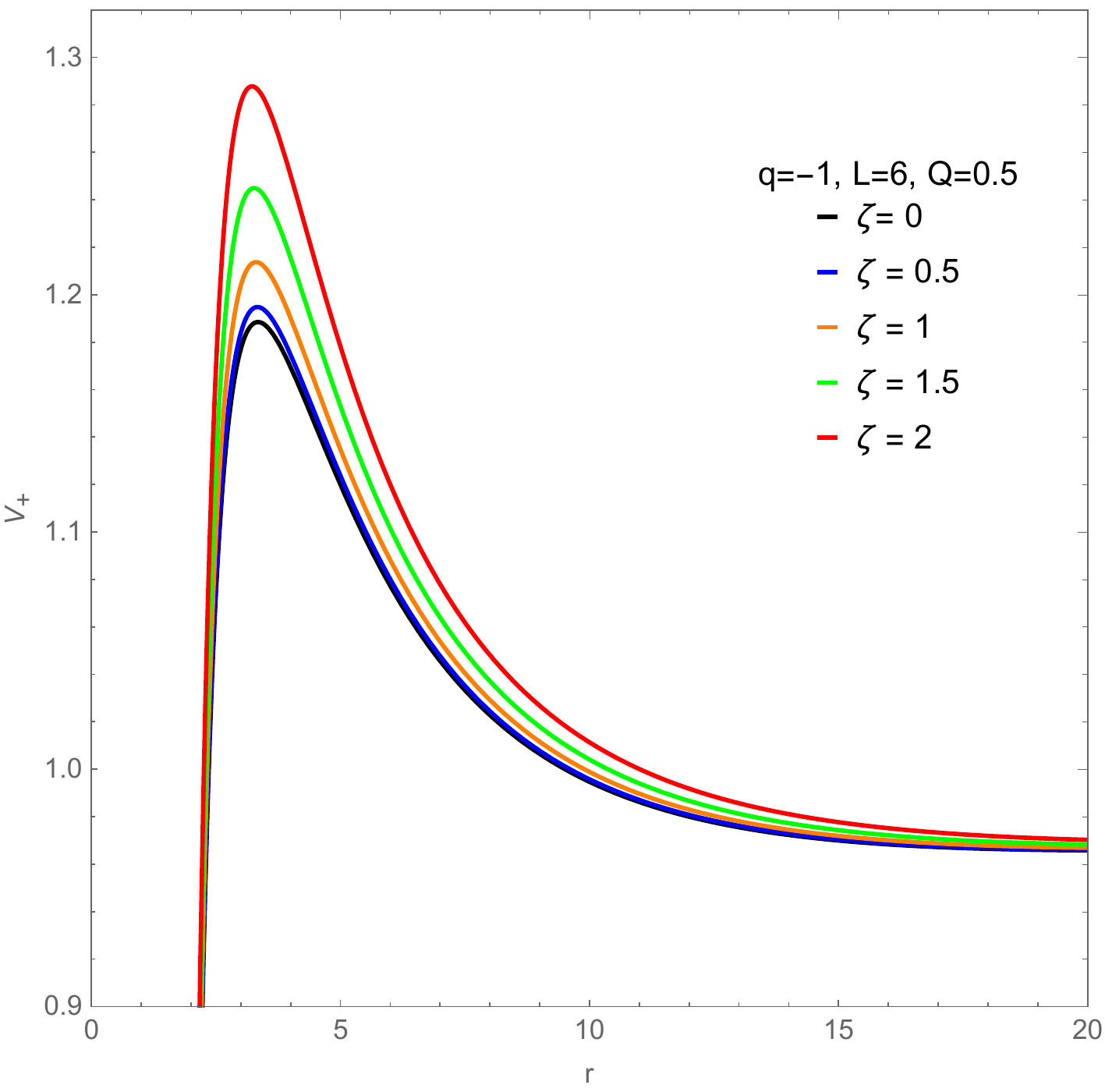}}

	\caption{The effective potential $V_+$ for different values of $\zeta$, with fixed $Q=0.5$ and $L=6$.}
	\label{fig_Vz2}
\end{figure*}

The relative magnitude between the effective potential $V_+$ and the specific energy $E$ of the test particle determines the particle's motion around the quantum-corrected BH. As shown in figure~\ref{fig_Vz}, the blue solid curve represents the effective potential $V_+$ as a function of $r$, while the red solid line indicates the specific energy $E$ of the test particle. When the blue curve lies below the red line (red region), the particle can move within the corresponding range of $r$; conversely, when the blue curve lies above the red line (blue region), the particle is forbidden in that $r$-interval. The points where the red and blue curves intersect represent turning points in the particle's motion. Subsequently, taking $Q=0.5$ and $L=6$ as an example, we show in figure~\ref{fig_Vz2} the variation of the effective potential $V_+$ for different values of the quantum parameter $\zeta$. A clear increasing trend of $V_+$ with increasing $\zeta$ can be observed.

\section{Energy extraction from the quantum-corrected BH}\label{section3}

\subsection{Generalized ergosphere}

Energy extraction from the quantum-corrected BH requires the existence of negative energy states for charged particles outside the event horizon $r_+$. This necessitates the condition $E=V_+ <0$~\cite{Baez:2024lhn}. The region supporting negative energy states is referred to as the ``generalized ergosphere", whose boundary $r_e$ is defined by $V_+ =0$. Combining this condition with eq.~\eqref{Eq:Vz} yields
\begin{equation}
	\begin{split}
		\frac{qQ}{r} + \sqrt{f(r)\left(1+\frac{L^2}{r^2}\right)}=0,
	\end{split}
\end{equation}
which imposes the constraint $qQ < 0$ on the charges of the BH and the test particle. Thus, the ergoregion boundary $r_e$ should satisfy
\begin{equation}
	\begin{split}
	r_e=\sqrt{\frac{q^2 Q^2-L^2 f(r_e)}{f(r_e)}}.\label{eq_re}
	\end{split}
\end{equation}

The ergoregion boundary $r_e$ obtained from the above equation must satisfy $r_e > r_+$. Furthermore, the sign of $L$ does not affect the boundary of the generalized ergoregion. Therefore, we will only consider cases with $L \geq 0$ in subsequent analysis.

\begin{figure*}[htbp]
	\centering
	\subfigure[$Q=0.5$ and $q=-6$]{\includegraphics[width=0.45\textwidth,width=3in, height=5in, keepaspectratio]{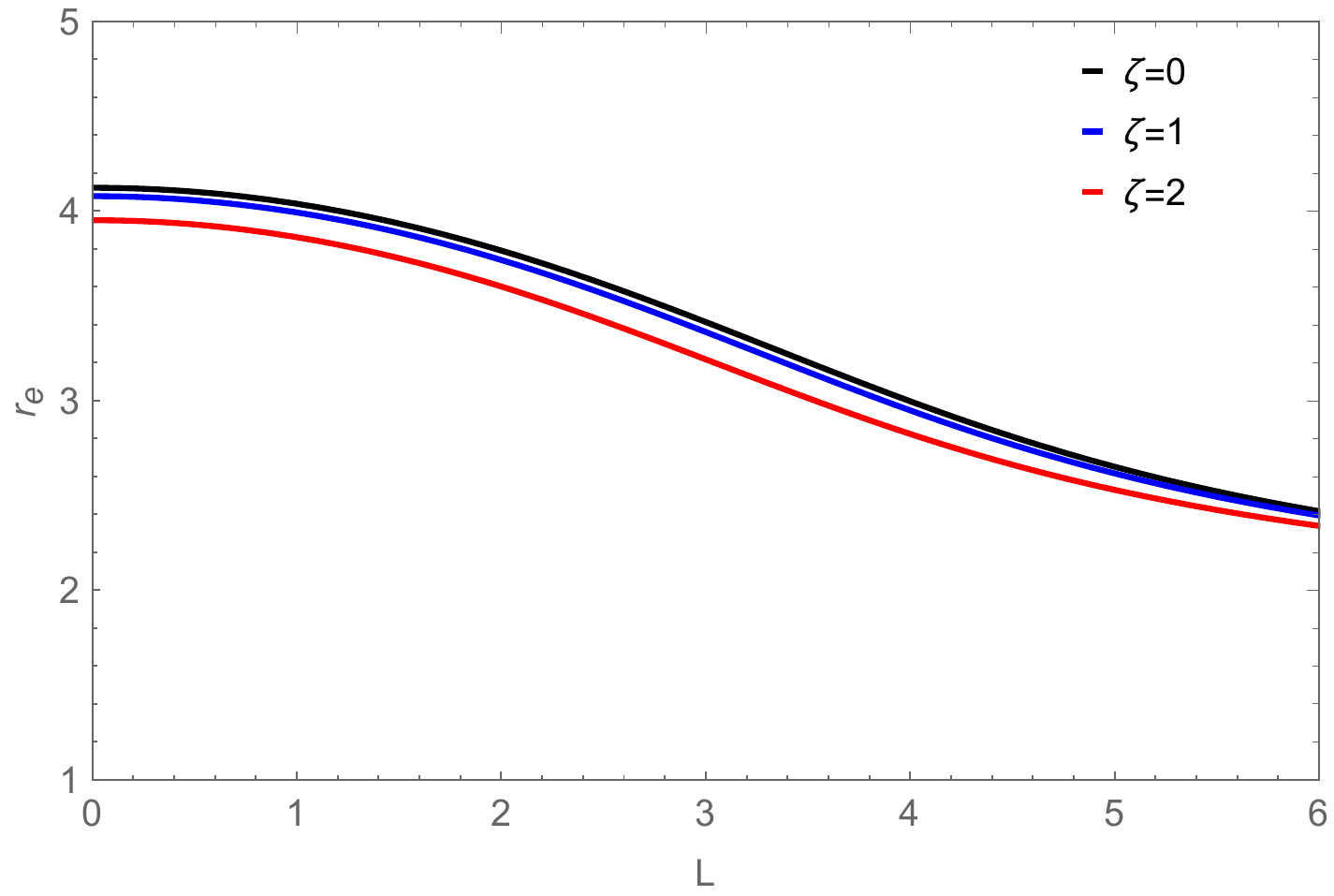}}
	%\hfill
	\subfigure[$Q=0.5$ and $L=6$]{\includegraphics[width=0.45\textwidth,width=3in, height=5in, keepaspectratio]{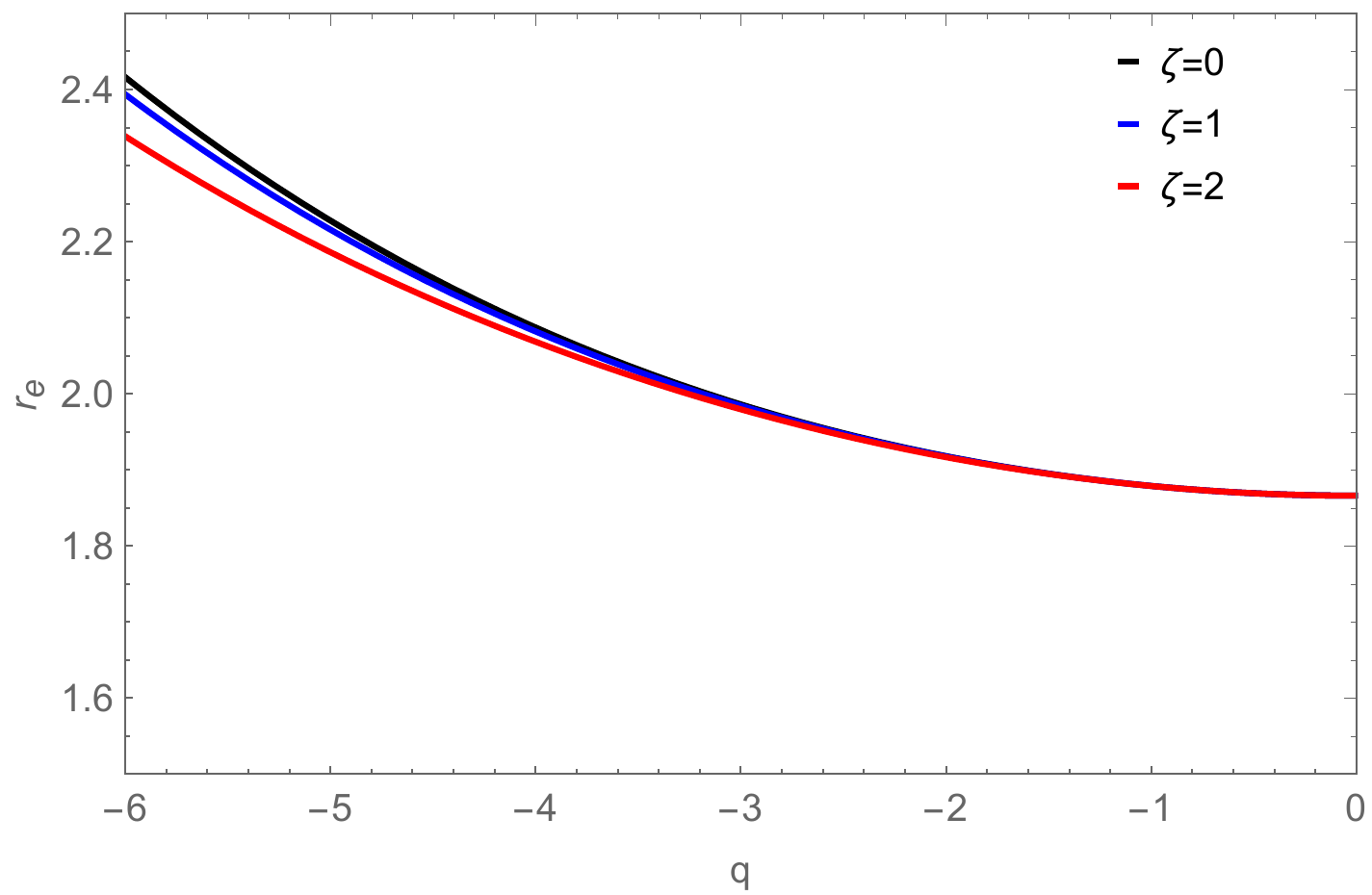}}

	\caption{Variation of the ergoregion boundary $r_e$ under different values of the parameters $L$, $q$, and $\zeta$.}
	\label{fig_re_zeta}
\end{figure*}

Under the fixed condition $Q=0.5$, figure~\ref{fig_re_zeta} shows the variation of the ergoregion boundary $r_e$ under different parameter values. It can be clearly observed that for charged particles with a negative charge ($q < 0$), a larger absolute charge $|q|$ and a smaller absolute angular momentum $|L|$ result in a larger $r_e$, and consequently a larger ergoregion. In contrast, the presence of $\zeta$ causes a reduction in $r_e$, which gradually decreases with increasing $\zeta$. This indicates that, under identical conditions, the ergoregion of the quantum-corrected BH is smaller than that of its classical counterpart.

\subsection{Energy extraction}

Next, we investigate the electric Penrose process for charged particles around this quantum-corrected BH. We consider an idealized astrophysical environment: a massive BH with no other massive objects in its vicinity, and its spacetime background remains stationary for a certain period of time. There is no plasma or accretion disk around the BH. A charged test particle, labeled particle 1, with mass much smaller than that of the BH, moves in the equatorial plane and is initially heading toward the BH from a distant location. At a certain point outside the event horizon, it splits into two fragments (denoted as particles 2 and 3), and the resulting fragments continue to move in the equatorial plane of the BH. At this moment, particle 2 is located within its corresponding generalized ergoregion, possessing negative energy and eventually falling into the BH. In contrast, particle 3 carries energy greater than that of particle 1. For simplicity, we assume the splitting point coincides with the turning points $r_t$ of all three particles.

In this process, both the particle charge $e$ and the 4-momentum $P^a$ are conserved, yielding
\begin{align}
	m_1 q_1 &= m_2 q_2 + m_3 q_3, \label{charge} \\
	P_1^a &= P_2^a + P_3^a. \label{4momentum}
\end{align}
Here, $m_i$, $q_i$, and $P_i^a$ denote the mass, specific charge, and 4-momentum of particle $i$ ($i=1,2,3$), respectively. And the masses of the particles satisfy $m_1 \geq m_2 + m_3$. Based on eq.~\eqref{4momentum}, the temporal and spatial components of the 4-momentum can be expressed as
\begin{align}
	m_1 E_1 &= m_2 E_2 + m_3 E_3, \label{energy} \\
	m_1 L_1 &= m_2 L_2 + m_3 L_3. \label{L_momentm}
\end{align}
Furthermore, from eq.~\eqref{Eq:Vz} we obtain the particle's energy at the turning point $r_t$ as
\begin{equation}
	E_i=\frac{q_{i} Q}{r_t}+\sqrt{f(r_t)\left(1+\frac{L_{i}^2}{r_t^2}\right)}\,, \quad i=1,2,3. \label{Ei}
\end{equation}
Then, the corresponding angular momentum can be solved as a function of the energy $E_i$ and the turning point $r_t$
\begin{equation}
	L_i=\pm\sqrt{\frac{r_t^2}{f(r_t)} \left[\left(E_i-\frac{q_{i} Q}{r_t}\right)^2-f(r_t)\right]}\,, \quad i=1,2,3. \label{angularmomentum}
\end{equation}
As mentioned previously, we only consider the case with $L \geq 0$, and thus select the positive sign for $L_i$ here. For convenience in discussion, we introduce the symbol $\mathcal{E}_{i}$, denoted as
\begin{equation}
	\mathcal{E}_{i} \equiv E_i-\frac{q_{i} Q}{r_t}, \quad i=1,2,3. \label{varE}
\end{equation}
Combining eqs.~\eqref{charge} and \eqref{energy}, we derive
\begin{equation}
	m_1 \mathcal{E}_{1} = m_2 \mathcal{E}_{2} + m_3 \mathcal{E}_{3}. \label{energy1}
\end{equation}
Substituting eqs.~\eqref{L_momentm} and \eqref{energy1} into eq.~\eqref{angularmomentum} and simplifying, we obtain
\begin{equation}
	m_1 \sqrt{\mathcal{E}_{1}^2 - f(r_t)} - m_2 \sqrt{\mathcal{E}_{2}^2 - f(r_t)} - m_3 \sqrt{\mathcal{E}_{3}^2 - f(r_t)} = 0.
\end{equation}
It should be emphasized that during the energy extraction process, particle 2 may not occupy a negative energy state following the initial setup of particle 1. Hence, we explicitly constrain the initial parameters of particle 2 to ensure it meets the negative energy requirement, namely,
\begin{equation}
	E_2=\frac{q_{2} Q}{r_t}+\sqrt{f(r_t)\left(1+\frac{L_{2}^2}{r_t^2}\right)}<0 \,. \label{E1}
\end{equation}
Therefore, using the corresponding conservation relations, we can obtain the energy and angular momentum of particles 1 and 3, namely
\begin{align}
		\mathcal{E}_1 &= \frac{\mathcal{E}_2 C_1 + \sigma \sqrt{C_3 \left(\mathcal{E}_2^2 - f(r_t)\right)}}{2 m_1 m_2}, \label{E0}\\
		\mathcal{E}_3 &= \frac{\mathcal{E}_2 C_2 + \sigma \sqrt{C_3 \left(\mathcal{E}_2^2 - f(r_t)\right)}}{2 m_2 m_3},\label{E2}\\
		L_1 &= \frac{L_2 C_1 + \sigma \sqrt{C_3 \left(L_2^2 + r_t^2\right)}}{2 m_1 m_2},\label{L0}\\
		L_3 &= \frac{L_2 C_2 + \sigma \sqrt{C_3 \left(L_2^2 + r_t^2\right)}}{2 m_2 m_3}, \label{L3}
\end{align}
where
\begin{align}
	C_1&=m_1^2 + m_2^2 - m_3^2,\label{c1}\\
	C_2&=m_1^2 - m_2^2 - m_3^2,\label{c2}\\
	C_3&=m_1^4 + \left(m_2^2 - m_3^2\right)^2 - 2 m_1^2 \left(m_2^2 + m_3^2\right),\\
	\sigma &=\pm 1.
	\label{L2}
\end{align}

The energy extraction efficiency $\eta$ throughout the entire process is
\begin{equation}
	\eta=\frac{m_3 E_3-m_1 E_1}{m_1 E_1}=-\frac{m_2 E_2}{m_1 E_1}.\label{eff1}
\end{equation}
From the aforementioned conservation relations, the energy and angular momentum of the particles can be determined. Consequently, the energy‑extraction efficiency for the electric Penrose process is obtained as~\cite{Baez:2024lhn}
\begin{equation}
	\eta = \frac{2m_2^2 E_2}{\dfrac{-2m_1 m_{2} q_1 Q}{r_t} - C_1 \left(E_{2} - \dfrac{q_2 Q}{r_t}\right) - \sigma\sqrt{C_3 \left[\left(E_{2} - \dfrac{q_2 Q}{r_t}\right)^2 - f(r_t)\right]}}.\label{xiaolv}
\end{equation}

\begin{figure*}[htbp]
	\centering
	\subfigure[$Q=0.4$]{\includegraphics[width=2.in, height=5in, keepaspectratio]{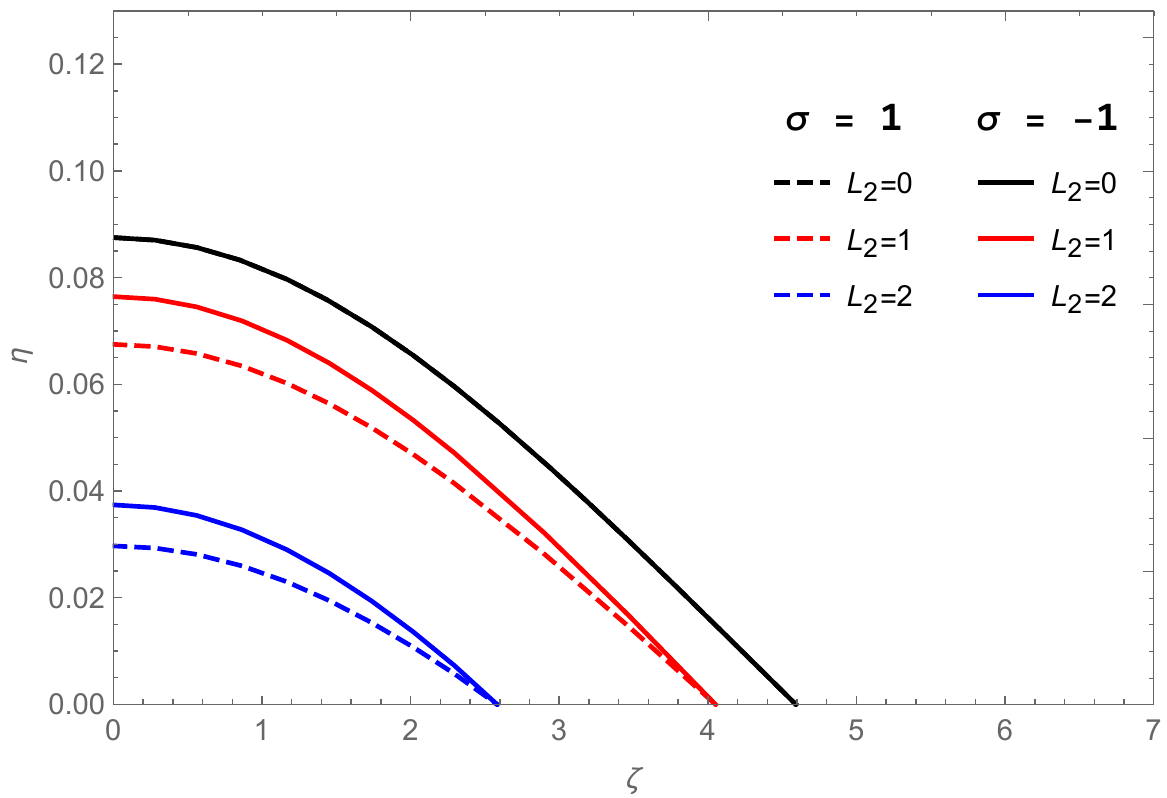}}
	\subfigure[$Q=0.5$]{\includegraphics[width=2.in, height=5in, keepaspectratio]{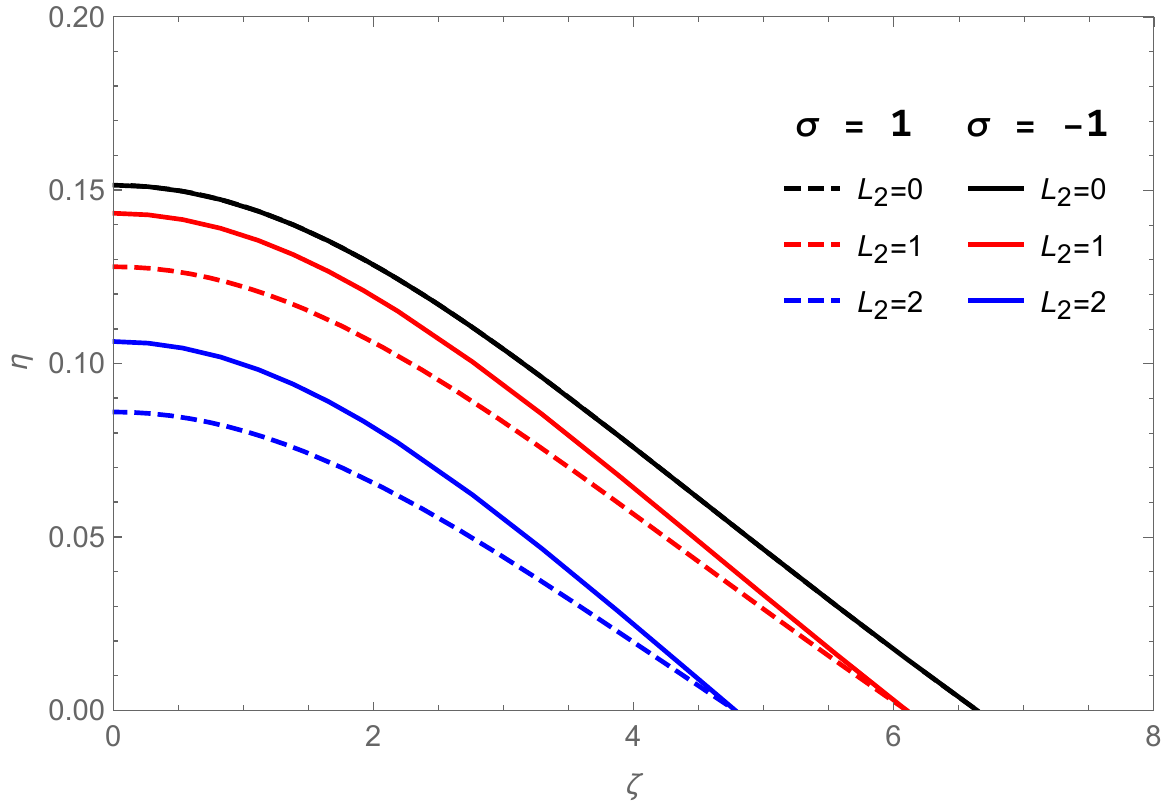}}
	\subfigure[$Q=0.6$]{\includegraphics[width=2.in, height=5in, keepaspectratio]{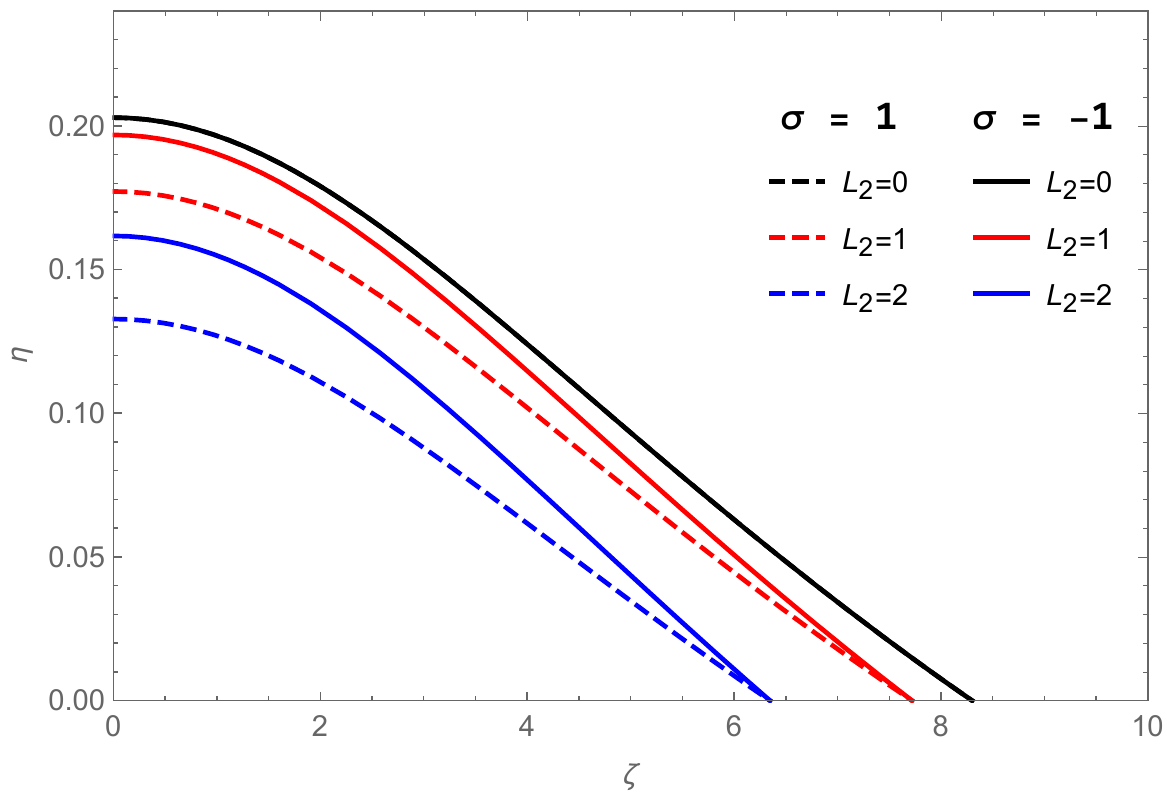}}

	\caption{Variation of the energy extraction efficiency $\eta$ with the quantum parameter $\zeta$ for different values of the BH charge $Q$ and angular momentum $L_2$. Dashed and solid curves correspond to $\sigma=1$ and $\sigma=-1$, respectively.}
	\label{fig_eta}
\end{figure*}

\begin{table*}
	\centering
	\setlength{\tabcolsep}{25pt}
	\begin{tabular}{|c| c| c| c|}
		\hline
		$i$ & $m_i$ & $q_i$ & $r_t$ \\
		\hline
		1 & $2$ & $3.3$ & $3$ \\
		2 & $0.9$ & $-6$ & $3$ \\
		3 & $1$ & $12$ & $3$ \\
		\hline
	\end{tabular}

	\caption{Fixed parameters (mass, specific charge, and turning point) for particles 1, 2, and 3. Unless otherwise specified, the BH charge is set to $Q = 0.5$.}
	\label{tab:1}
\end{table*}
Note that the efficiency $\eta$ is a function of ($Q$, $\zeta$, $L_2$, $m_i$, $q_i$). For fixed particle mass, charge, and turning point parameters (see Table~\ref{tab:1}), figure~\ref{fig_eta} shows the variation of $\eta$ with the quantum parameter $\zeta$ for different values of the BH charge $Q$ and angular momentum $L_2$. Equation~\eqref{eq_re} shows that different values of $Q$ yield different ergoregion boundaries $r_e$, which affects the choice of the particle's turning point $r_t$. To ensure that $r_t=3$ in table~\ref{tab:1} is satisfied, we select three cases with $Q=0.4, 0.5, 0.6$ here. In these figures, dashed and solid curves correspond to $\sigma=1$ and $\sigma=-1$, respectively. For different values of angular momentum $L_2$, $\zeta$ consistently reduces the energy extraction efficiency, while an increase in $Q$ leads to an increase in the extraction efficiency under the same conditions. Furthermore, when $L_2 = 0$, $\eta$ is the same for $\sigma = \pm 1$, while for $L_2 > 0$, $\sigma = -1$ yields a slightly higher $\eta$ than $\sigma = 1$.

\begin{figure}[htbp]
	\centering
	{\includegraphics[width=3.in, height=5in, keepaspectratio]{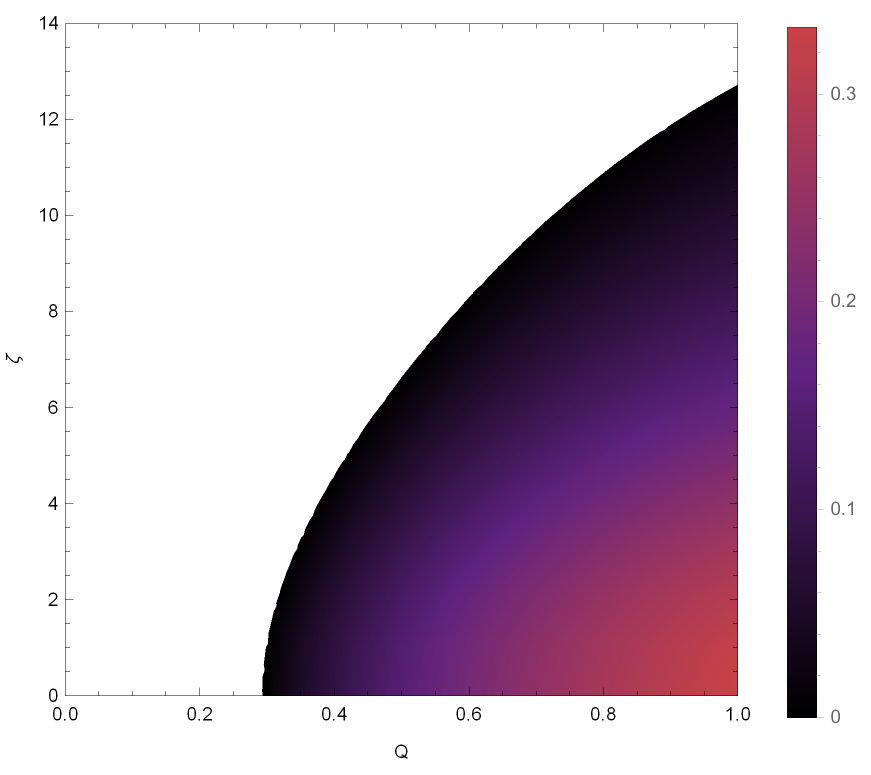}}

	\caption{Variation of $\eta$ in the parameter space $(Q, \zeta)$ with $L_2 = 0$.}
	\label{fig_eta1}
\end{figure}

In figure~\ref{fig_eta}, we also observe that at certain extreme values of $\zeta$, the energy extraction efficiency drops to zero (the critical $\zeta_c$ values are the same for both $\sigma=1$ and $\sigma=-1$). This occurs because increasing $\zeta$ shrinks the ergoregion boundary $r_e$; in extreme cases, the turning point $r_t$ satisfies $r_t > r_e$, making energy extraction impossible and thus driving the efficiency to zero. Subsequently, using the $m_i$, $q_i$, and $r_t$ from table~\ref{tab:1} and taking $L_2 = 0$ as an example, figure~\ref{fig_eta1} illustrates the density plot of $\eta$ in the parameter space spanned by $(Q, \zeta)$. The white region in the figure represents the area where energy extraction is impossible. It can be seen that for $r_t = 3$, both the quantum-corrected BH and the RN BH allow energy extraction only when $Q$ takes relatively large values. From the plot, the same conclusion as above can be drawn: the energy extraction efficiency of this type of quantum‑corrected BH is always lower than that of the classical RN BH, and there exists a critical $\zeta_c$ at which the efficiency drops to zero. From eq.~\eqref{xiaolv}, the condition $\eta = 0$ yields $E_2 = 0$, and the critical quantum parameter $\zeta_c$ depends on $Q$, $L_2$, $q_2$, $r_t$ as follows:
\begin{equation}
	\zeta_c = \frac{r_t^2 \sqrt{Q^2 q_2^2 r_t^2 - \left(r_t^2 - 2r_t + Q^2\right)\left(r_t^2 + L_2^2\right)}}{\left(r_t^2 - 2r_t + Q^2\right) \sqrt{r_t^2 + L_2^2}}.\label{zeta_c}
\end{equation}
The quantum parameters in the following are all below $\zeta_c$. To simplify the subsequent analysis, we hereafter consider only the case where the angular momentum of particle 2 is zero, i.e., $L_2 = 0$.

\section{Particle motion in a general electric Penrose process}\label{section4}

In the above discussion, we have only considered the energy extraction efficiency when particle 1 splits into two fragments (particles 2 and 3) at the turning point, without addressing the subsequent motion of the resulting particles. Whether particle 3 --- produced from the splitting of particle 1 that started from a distant observer --- can carry the gained energy back to the observer determines whether the extracted energy is utilizable. In Ref.~\cite{Vertogradov:2025wgg}, a rough proof was given for the case in which particle 3 with zero angular momentum ($L_3 = 0$) can escape the BH. Here, we consider the escape of particle 3 under more relaxed conditions.

The turning point $r_t$ of particle 1 (which comes from a distant observer) must lie outside the peak of its effective potential, i.e., $r_t > r_m$, with $r_m$ being the radial coordinate of the extremum shown in figure~\ref{fig_Vz}. In this scenario, if no splitting occurs, particle 1 would move outward from $r_t$. Here, the effective potential of particle 1 is monotonically decreasing in a small region immediately to the right of $r_t$, and its derivative $V_{1}'$ satisfies
\begin{equation}
	\begin{split}
		V_{1}' &= \frac{-2 q_1 Q \sqrt{\left(L_1^2+r_t^2\right) f(r_t)} + L_1^2\left[r_t f'(r_t)-2 f(r_t)\right] + r_t^3 f'(r_t)}{2 r_t^2 \sqrt{\left(L_1^2 +r_t^2\right) f(r_t)}} < 0.\label{Eq:V1}
	\end{split}
\end{equation}
The derivative of the effective potential $V_3 '$ for particle 3 after the splitting is
\begin{equation}
	\begin{split}
		V_{3}' = \frac{-2 q_3 Q \sqrt{\left(L_3^2 + r_t^2\right) f(r_t)} \;+\; L_3^2\left[r_t f'(r_t)-2 f(r_t)\right] \;+\; r_t^3 f'(r_t)}{2 r_t^2 \sqrt{\left(L_3^2 + r_t^2\right) f(r_t)}}.\label{Eq:V3}
	\end{split}
\end{equation}
We first consider the case $L_2 = 0$. Combining eqs.~\eqref{charge} and \eqref{L_momentm}, the term that determines the sign of $V_{3}'$ can be written as
\begin{align}
	V_{3}' \propto &-2 q_3 Q \sqrt{\left(M_1 ^2 L_1^2 + r_t^2\right) f(r_t)} \;+\; M_1 ^2 L_1^2\left[r_t f'(r_t)-2 f(r_t)\right]+\; r_t^3 f'(r_t)\;,\label{Eq:V33}
\end{align}
where
\begin{equation}
	\begin{split}
		M_1=m_1/m_3.
	\end{split}
\end{equation}
Note that the mass relation between the particles requires $M_1 > 1$, and that $f'(r) > 0$ always holds outside the horizon. In addition, since $q_2 < 0$, combining with eq.~\eqref{charge} we have
\begin{equation}
	\begin{split}
		q_3>M_1 q_1>0.\label{eq_qm}
	\end{split}
\end{equation}

If the expression $G_{1}(r) \equiv r f'(r) - 2 f(r)$ evaluated at the turning point $r_t$ is less than or equal to zero, a direct comparison between eqs.~\eqref{Eq:V33} and \eqref{Eq:V1} shows that eq.~\eqref{Eq:V33} must be negative at the same turning point $r_t$. This demonstrates that the condition $V_{3}' < 0$ holds for particle 3; therefore, starting from the turning point, particle 3 will move away from the BH.

\begin{figure}[htbp]
	\centering
	\includegraphics[width=3.2in, height=5in, keepaspectratio]{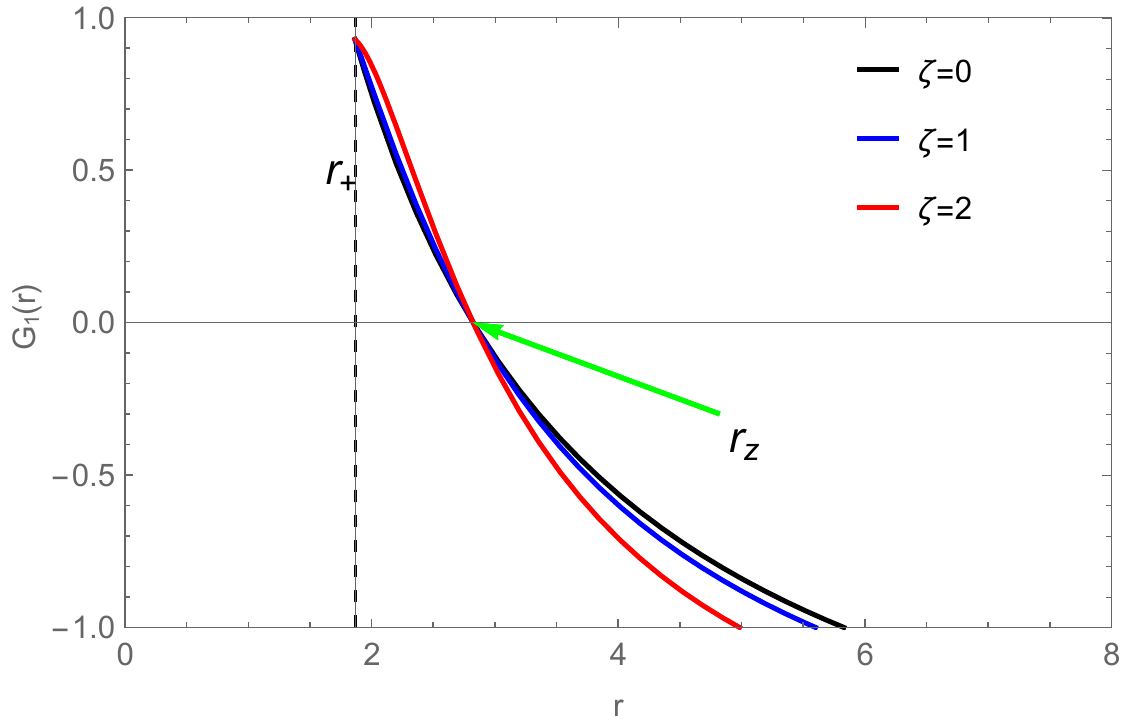}

	\caption{Behavior of $G_{1}(r)$ for different values of $\zeta$ with $Q=0.5$.}
	\label{fig_Gr}
\end{figure}

The function $G_{1}(r)$ depends solely on the spacetime geometry of the BH. Substituting eq.~\eqref{metric1} into $G_{1}(r)$, we obtain
\begin{equation}
	\begin{split}
		G_{1}(r)=-\frac{2 \left(2 Q^2 + r (-3 M + r)\right) \left(r ^4 + 2 Q^2 \zeta^2 - 4 M r \zeta^2 + 2 r ^2 \zeta^2\right)}{r ^6}.
	\end{split}
\end{equation}
The solution of $G_{1}(r)=0$ located outside the quantum-corrected BH horizon $r_+$ is given by
\begin{equation}
	\begin{split}
		r_z = \frac{3M}{2} + \frac{1}{2}\sqrt{9M^2 - 8Q^2}. \label{eq_rz}
	\end{split}
\end{equation}
Note that the root $r_z$ of $G_{1}(r)$ outside the quantum-corrected BH horizon is independent of the quantum parameter $\zeta$. It is straightforward to show that $G_{1}(r) \leq 0$ in the region where $r_t \geq r_z$, while $G_{1}(r) > 0$ for $r_t < r_z$. Taking $Q = 0.5$ as an example, we illustrate the behavior of $G_{1}(r)$ as a function of $r$ in figure~\ref{fig_Gr}. Consequently, determining the sign of $V_{3}'$ requires further verification for the case $G_{1}(r) > 0$.

Based on the established conditions, we have proven that even when $G_{1}(r) > 0$, the derivative of the effective potential for particle 3, $V_{3}'$, remains negative at the same turning point. The detailed proof is provided in appendix~\ref{appendix}. Consequently, we can conclude that for the simplified scenario with zero angular momentum for particle 2 ($L_2 = 0$), particle 3 --- which gains more energy from the splitting of particle 1 at its turning point --- can always escape the BH and carry the extracted energy away. Moreover, we discussed the escape of particle 3 under the relaxed conditions $L_2 \neq 0$ in appendix~\ref{appendix1}. The results show that particle 3 can escape for $L_2 > 0$ and $\sigma = 1$, as well as for $L_2 < 0$ and $\sigma = -1$. However, for the complementary cases --- $L_2 > 0$ with $\sigma = -1$, and $L_2 < 0$ with $\sigma = 1$ --- a proof that particle 3 always escapes the BH has not yet been obtained, and a detailed analysis based on the specific angular momentum values is necessary. It is worth noting that throughout the proof, we only imposed minimal requirements on the metric function: $f'(r) > 0$ and $f(r) > 0$ outside the event horizon. Beyond these, no specific spacetime restrictions were applied. Therefore, the conclusions obtained under this framework are applicable to particle motion in the electric Penrose process across a wide range of theoretical models.

\begin{figure*}[htbp]
	\centering
	\subfigure[$\zeta=0$]{\includegraphics[width=0.329\textwidth,height=4.5cm, keepaspectratio]{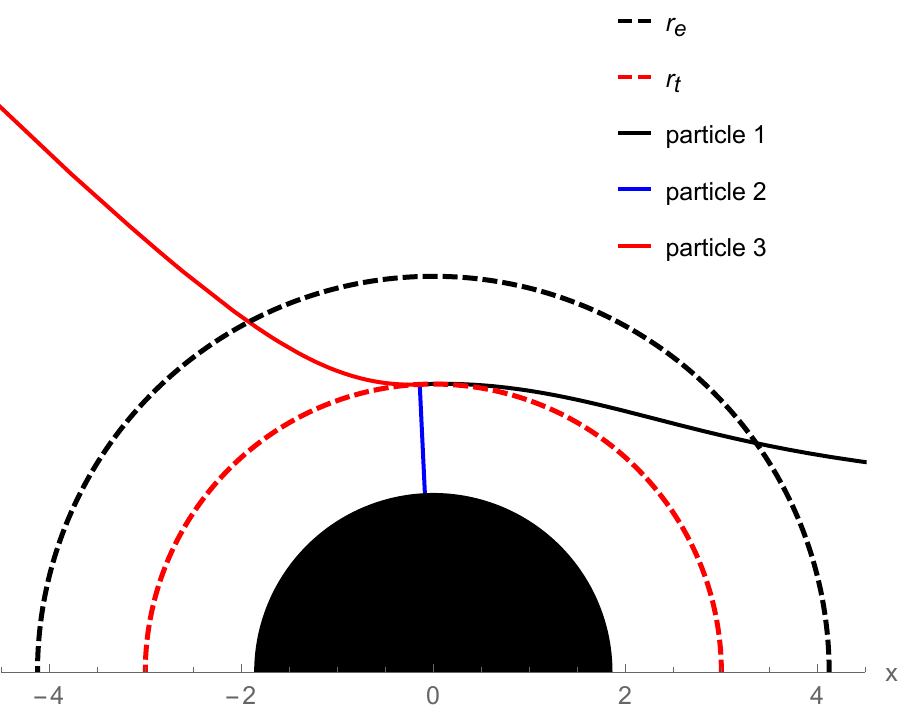}}
	\subfigure[$\zeta=1$]{\includegraphics[width=0.329\textwidth,height=4.5cm, keepaspectratio]{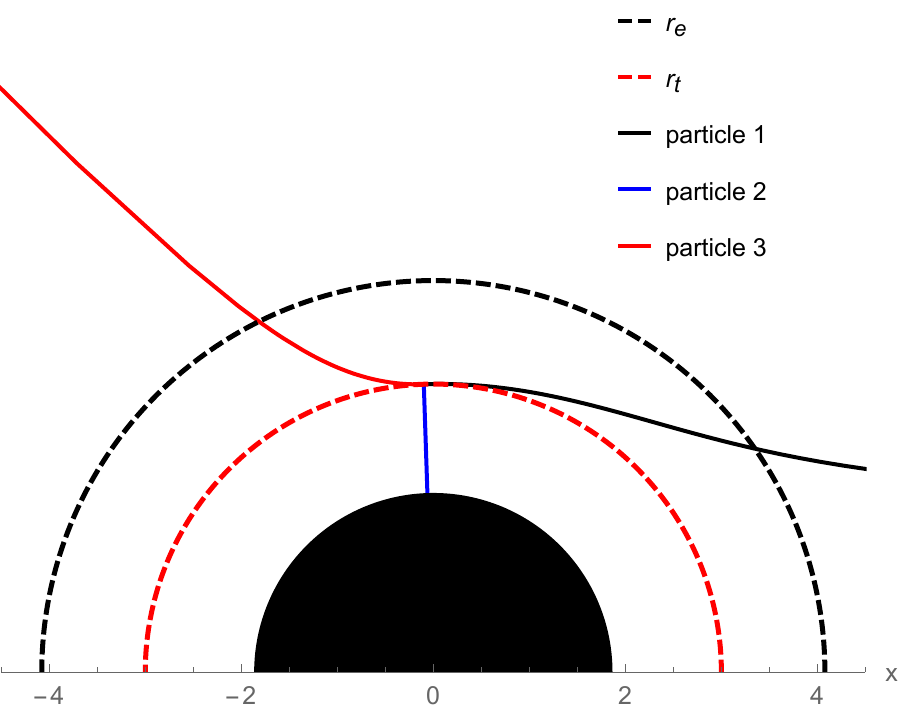}}
	\subfigure[$\zeta=2$]{\includegraphics[width=0.329\textwidth,height=4.5cm, keepaspectratio]{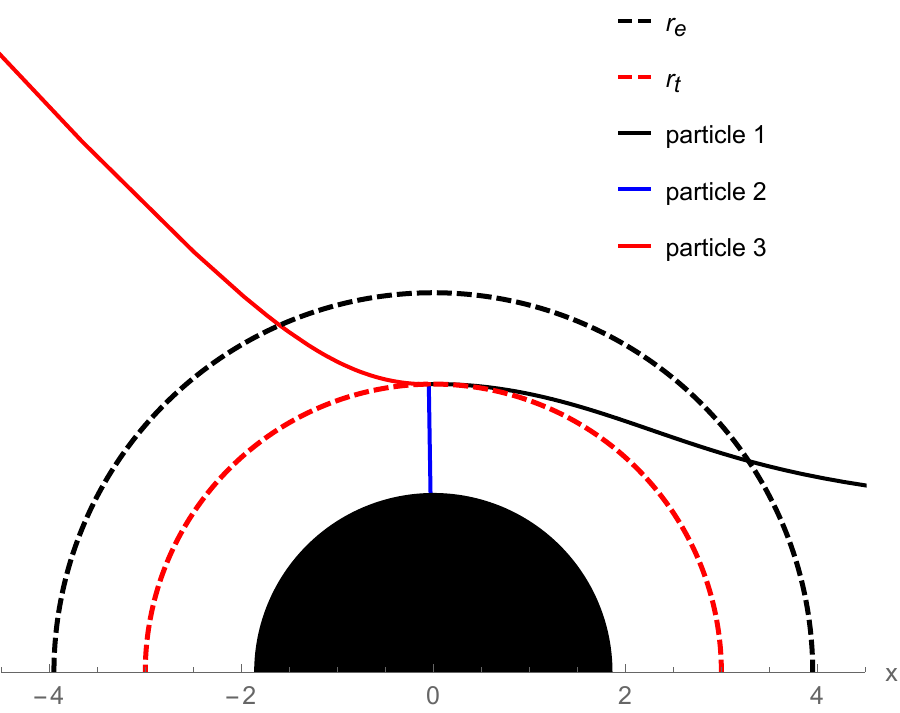}}

	\caption{Trajectories of particles 1, 2, and 3 for different values of $\zeta$ at the turning point $r_t=3$. The black semi-disk represents the BH. The black curve shows the trajectory of particle 1, while the blue and red solid curves correspond to the trajectories of particles 2 and 3, respectively.}
	\label{fig_guiji}
\end{figure*}

Subsequently, using the initial parameter values listed in Table~\ref{tab:1} and $Q = 0.5$, we present in figure~\ref{fig_guiji} a partial trajectory that illustrates the motion following the split of particle 1 into particles 2 and 3 at the turning point $r_t = 3$. In the figure, the black dashed line represents the ergoregion boundary $r_e$, the red dashed line corresponds to a circle with radius equal to the turning point $r_t$, and the black semi-disk denotes the BH. Additionally, the black, blue, and red solid curves depict the trajectories of particle 1, particle 2, and particle 3, respectively. It can be observed that particle 3, produced from the splitting of particle 1 originating from a distant observer, gradually moves away from the BH. Comparing the particle trajectories for different values of $\zeta$, we find that $\zeta$ exerts a slight influence on the particle's motion.

\section{Particle motion for a special electric Penrose process}\label{section5}

As discussed above, when the effective potential of a particle exhibits a peak (similar to figure~\ref{fig_Vz}), the region where the particle can possibly move includes not only the exterior of the peak ($r > r_m$) but also a portion inside the peak ($r_+ < r < r_m$), even though a particle moving inside this region cannot escape the BH. If we consider the electric Penrose process in which the particle's turning point lies between $r_+$ and $r_m$, then particle 1 cannot escape the BH from the outset. We further examine how the fragments produced by its splitting move. To distinguish this from the general electric Penrose process discussed above, we denote the Penrose process with a turning point located between $r_+$ and $r_m$ as the special electric Penrose process. The energy and angular momentum involved in this special electric Penrose process also satisfy the equations given in section~\ref{section3}. In the following, we will examine the motion of each particle involved in this process.

\begin{figure*}[htbp]
	\centering
	\subfigure[particle 1]{\includegraphics[width=0.329\textwidth,height=4.5cm,keepaspectratio]{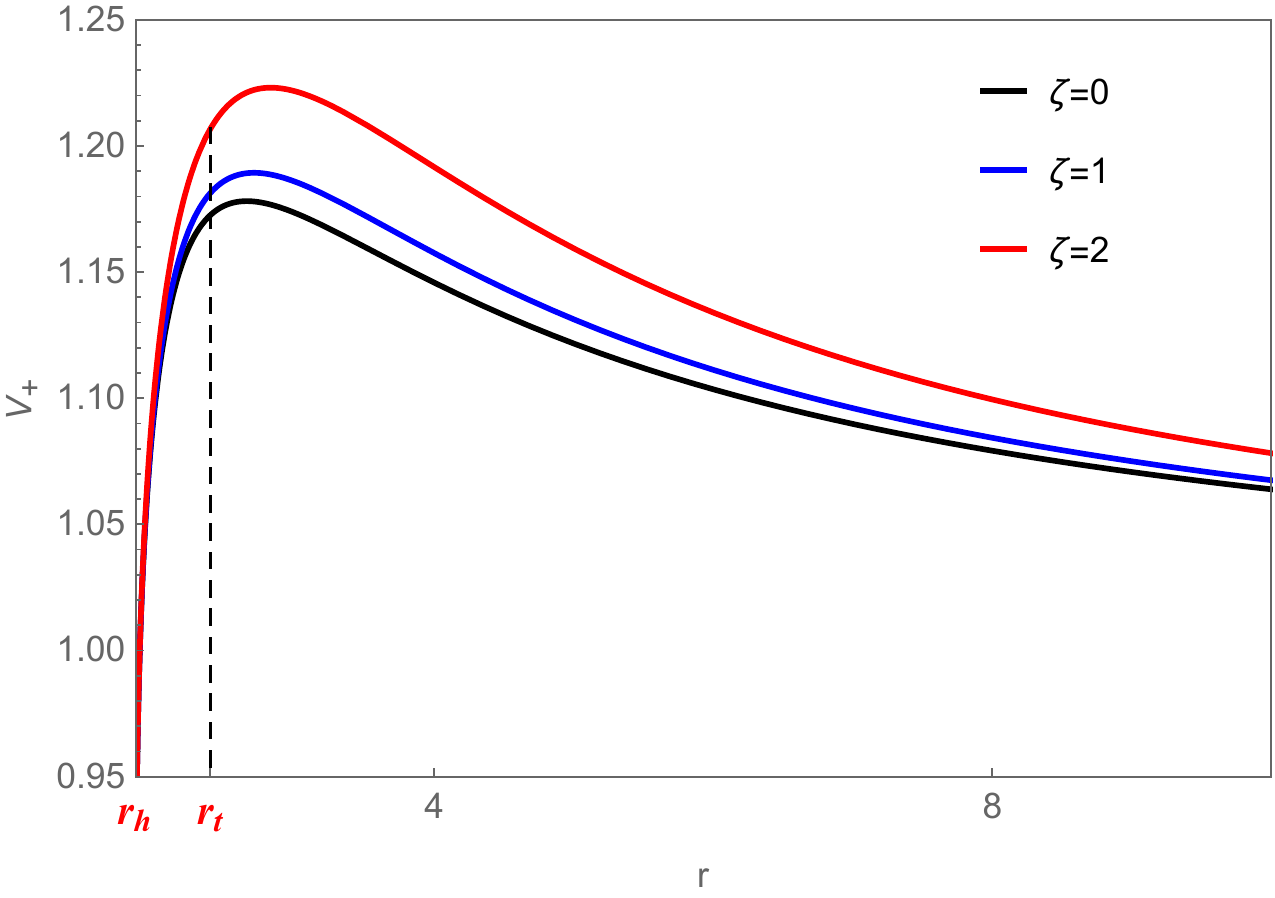}}
	\subfigure[particle 2]{\includegraphics[width=0.329\textwidth,height=4.5cm,keepaspectratio]{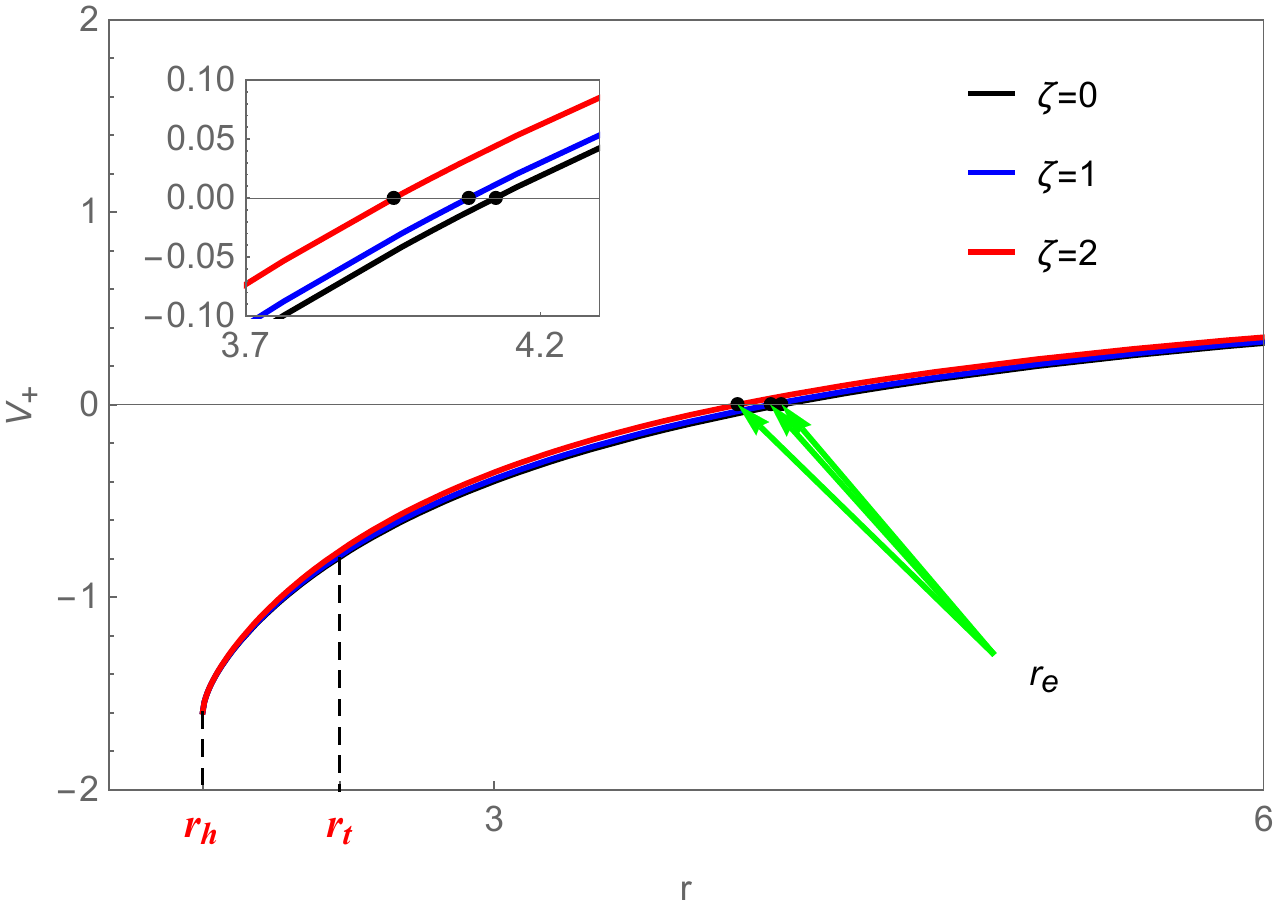}}
	\subfigure[particle 3]{\includegraphics[width=0.329\textwidth,height=4.5cm, keepaspectratio]{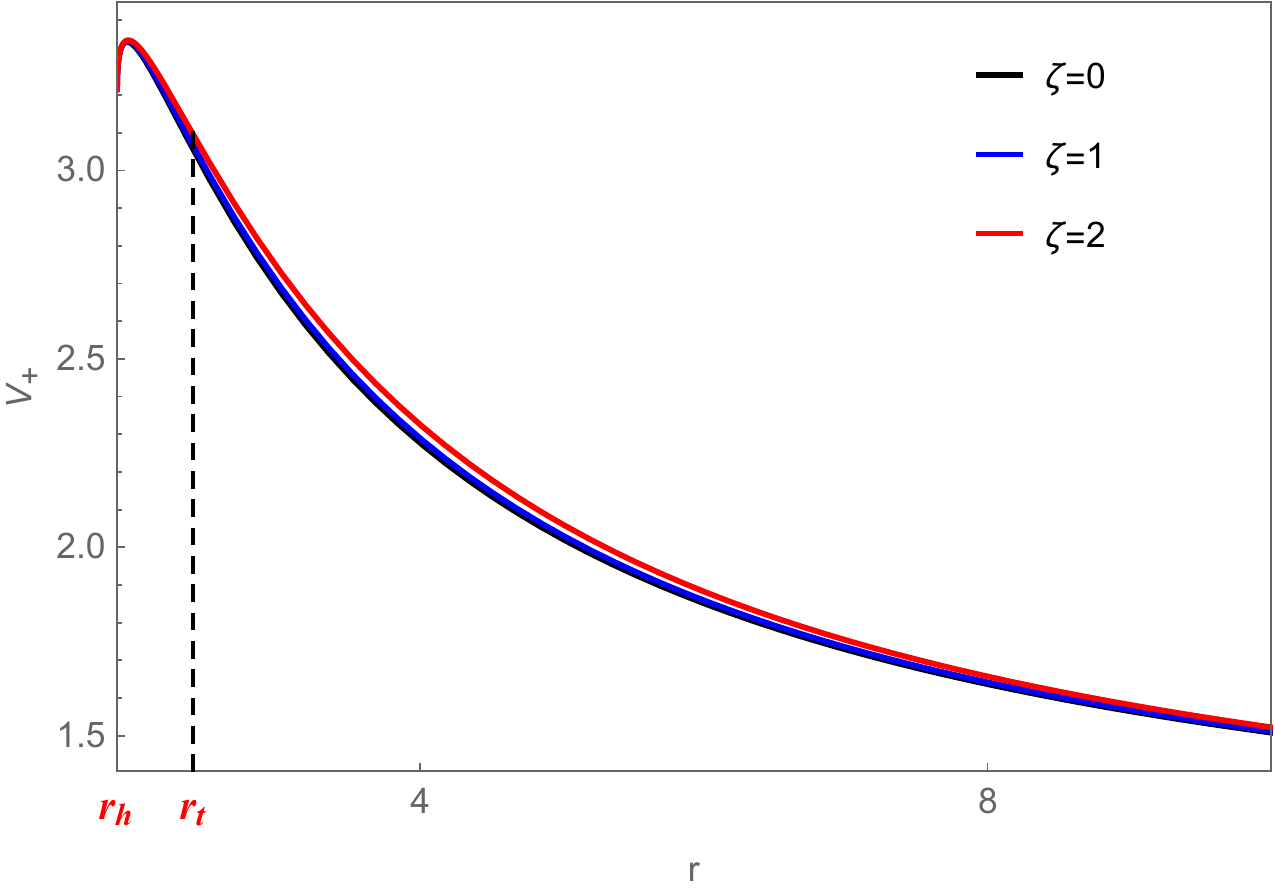}}

	\caption{Effective potentials $V_+$ of particles 1, 2, and 3 for different values of the quantum parameter $\zeta$, at the turning point $r_t=2.4$.}
	\label{fig_9}
\end{figure*}

\begin{figure*}[htbp]
	\centering
	\subfigure[$\zeta=0$]{\includegraphics[width=0.329\textwidth,height=4.5cm,keepaspectratio]{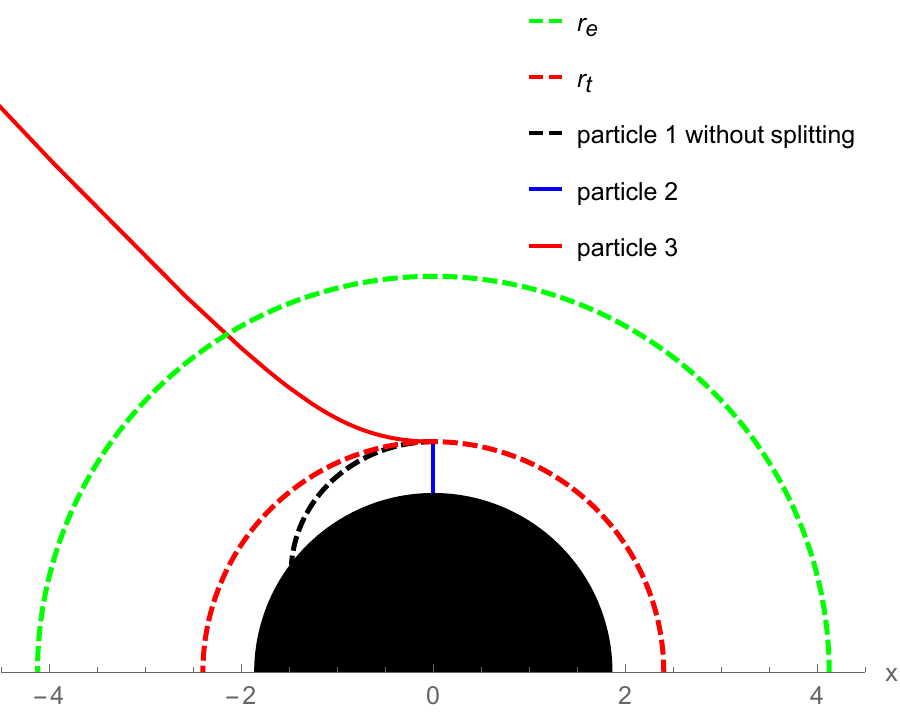}}
	\subfigure[$\zeta=1$]{\includegraphics[width=0.329\textwidth,height=4.5cm,keepaspectratio]{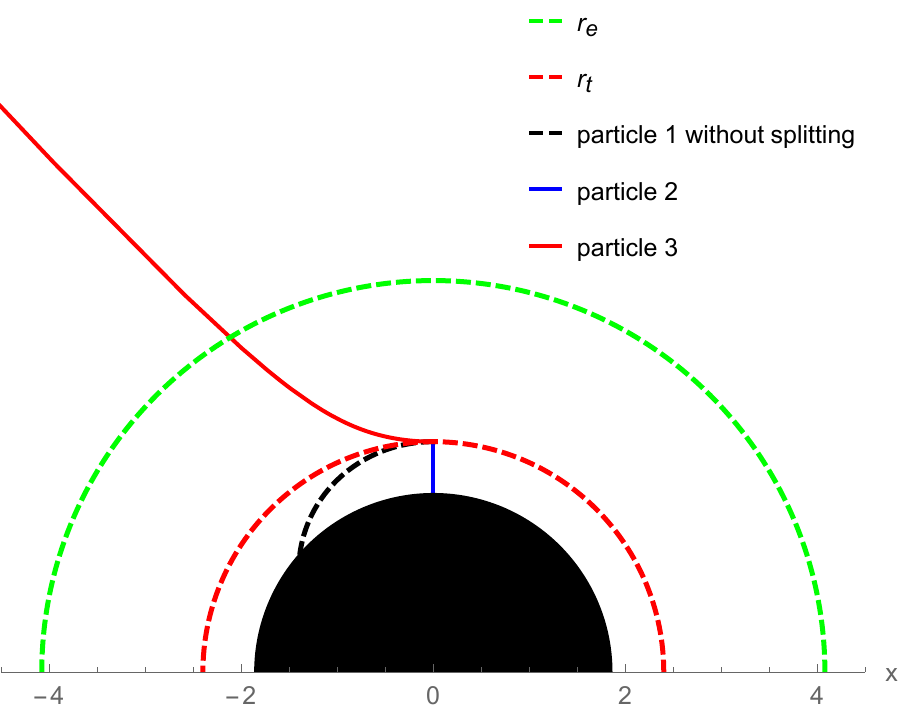}}
	\subfigure[$\zeta=2$]{\includegraphics[width=0.329\textwidth,height=4.5cm,keepaspectratio]{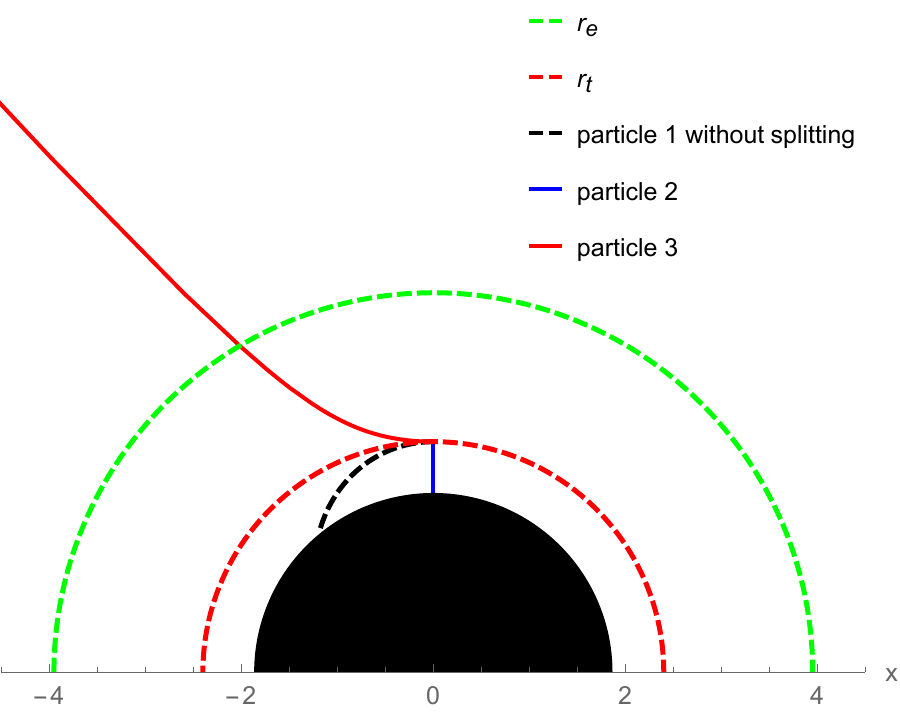}}

	\caption{Trajectories of particles 1, 2, and 3 for different values of $\zeta$ at the turning point $r_t=2.4$. The black semi-disk represents the BH. The black dashed curve shows the trajectory of particle 1 without splitting, while the blue and red solid curves correspond to the trajectories of particles 2 and 3, respectively.}
	\label{fig_guiji1}
\end{figure*}

We also adopt the mass $m_i$ and specific charge $q_i$ parameters for each particle from table~\ref{tab:1} and $Q = 0.5$. In contrast to the previous discussion, we now choose a turning point closer to the BH horizon. When we set the particle turning point to $r_t=2.4$, the effective potential $V_+$ for particles 1, 2, and 3 under different values of $\zeta$ is shown in figure~\ref{fig_9}. It can be observed in figure~\ref{fig_9} that the presence of the quantum parameter $\zeta$ enhances the peak values of the effective potential $V_+$ for both particles 1 and 3, which is consistent with our earlier analysis. The motion of each particle in this scenario is shown in figure~\ref{fig_guiji1}.

In figure~\ref{fig_guiji1}, the green and red dashed lines represent the ergoregion boundary $r_e$ and the circle of radius $r_t$, respectively. If particle 1, located inside the ergosphere, were not to split at its turning point $r_t$, its subsequent trajectory would follow the black dashed curve shown in figure~\ref{fig_guiji1}. In the case where particle 1 splits at the turning point into particles 2 and 3, their trajectories are given by the blue and red solid lines, respectively. It is evident that if particle 1 does not split, it will inevitably fall into the BH and cannot return to a distant observer. However, the fragment produced from its splitting can escape the BH. Furthermore, the overall variation of trajectories across different values of $\zeta$ remains similar, indicating that the influence of $\zeta$ on particle motion is relatively weak.
\begin{figure*}[htbp]
	\centering
	\subfigure[particle 1]{\includegraphics[width=0.329\textwidth,height=4.5cm,keepaspectratio]{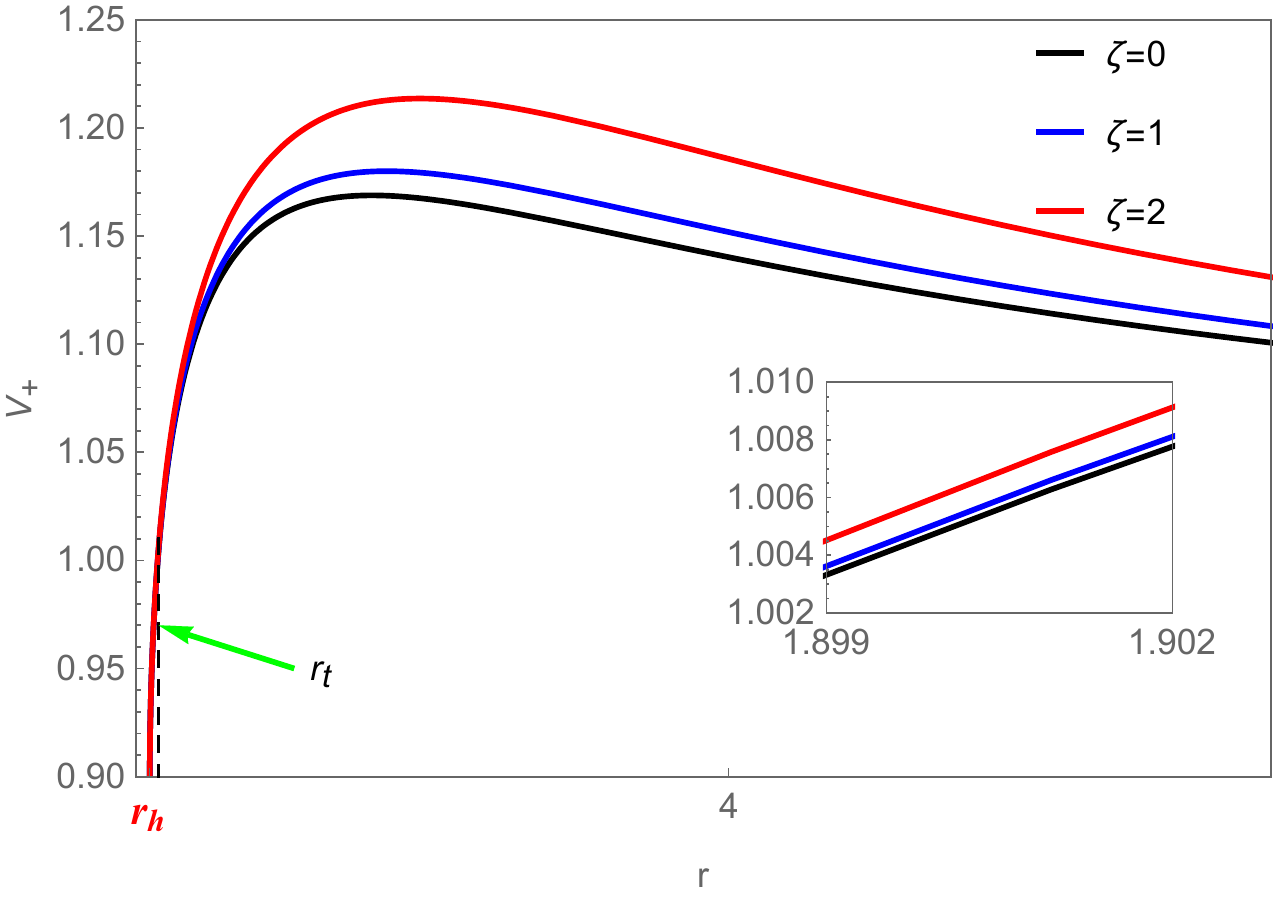}}
	\subfigure[particle 2]{\includegraphics[width=0.329\textwidth,height=4.5cm,keepaspectratio]{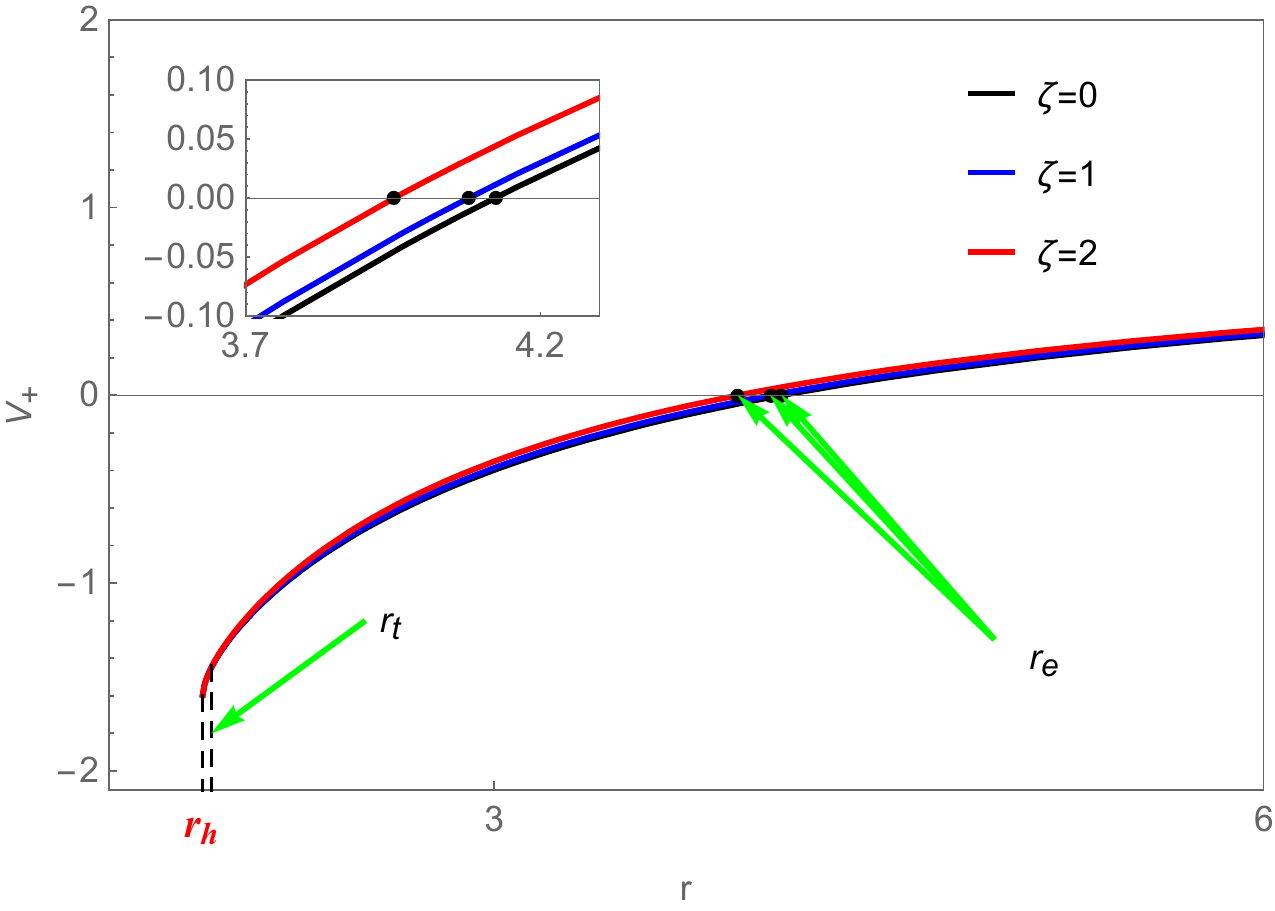}}
	\subfigure[particle 3]{\includegraphics[width=0.329\textwidth,height=4.5cm,keepaspectratio]{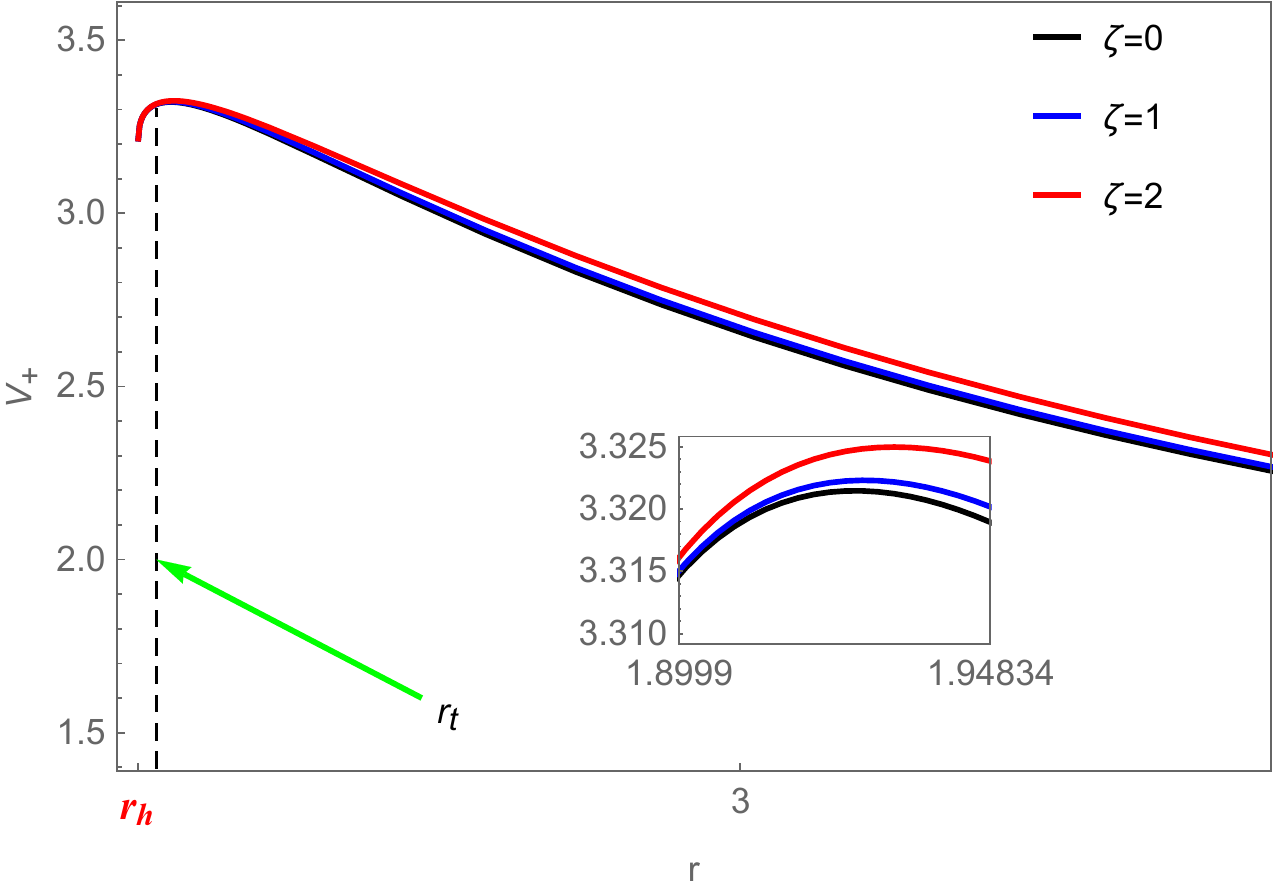}}

	\caption{Effective potentials $V_+$ of particles 1, 2, and 3 for different values of the quantum parameter $\zeta$, at the turning point $r_t=1.9$.}
	\label{fig_10}
\end{figure*}

\begin{figure*}[htbp]
	\centering
	\includegraphics[width=0.99\textwidth, height=8in, keepaspectratio]{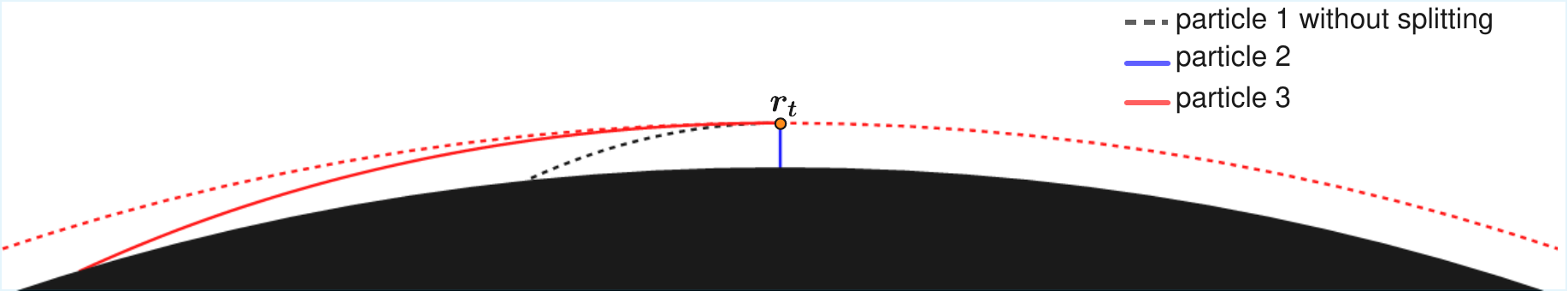}

	\caption{Partial trajectories of the particles at the turning point $r_t = 1.9$. The red dashed line denotes a circle of radius $r_t$, and the black dashed curve shows the trajectory of particle 1 without splitting. The blue and red solid curves represent the trajectories of particles 2 and 3, respectively. The black region indicates a partial cross-section of the BH.}
	\label{fig_no}
\end{figure*}

While keeping the mass and charge parameters unchanged, as the turning point $r_t$ moves closer to the BH horizon, the behavior of particle 3 exhibits a significant deviation from the cases described above. The effective potentials of all particles for $r_t = 1.9$ are shown in figure~\ref{fig_10}. Since the results for $\zeta = 1$ and $\zeta = 2$ closely resemble those for $\zeta = 0$, we use $\zeta = 0$ as a representative case and plot the corresponding particle trajectories in figure~\ref{fig_no}. The curve styles in these figures share the same meaning as those in figure~\ref{fig_guiji1}. The results demonstrate that when the turning point $r_t$ is too close to the horizon $r_+$, particle 3, which moves away from the BH in figure~\ref{fig_guiji1}, becomes unable to escape the BH in this scenario. In summary, for a particle 1 that may eventually fall into the BH, the fragment particle 3 produced via the electric Penrose process can exhibit two distinct dynamical outcomes: escaping from the BH (figure~\ref{fig_guiji1}) or falling into it (figure~\ref{fig_no}).

In this special electric Penrose process, the location of the turning point $r_t$ largely determines the fate of particle 3. In most cases, the influence of quantum parameter $\zeta$ on the particle trajectories is weak. However, under specific conditions, the $\zeta$ can also induce a qualitative change in the particle's behavior. We observe that for the quantum-corrected BH in figures~\ref{fig_9} and \ref{fig_10}, increasing the $\zeta$ shifts the radial coordinate $r_m$ corresponding to the potential peaks of particles 1 and 3 outward. This implies that the radial coordinate $r_m^{(3)}$ of the extremum in the effective potential of particle 3 in the RN case is smaller than its counterpart $R_m^{(3)}$ in the quantum-corrected BH case. When the turning point $r_t$ of particle 1 lies within the interval $r_m^{(3)} < r_t < R_m^{(3)}$, the motion of particle 3 may differ between the RN BH and its quantum-corrected counterpart.

\begin{figure*}[htbp]
	\centering
	\subfigure[particle 1]{\includegraphics[width=0.329\textwidth,height=4.5cm,keepaspectratio]{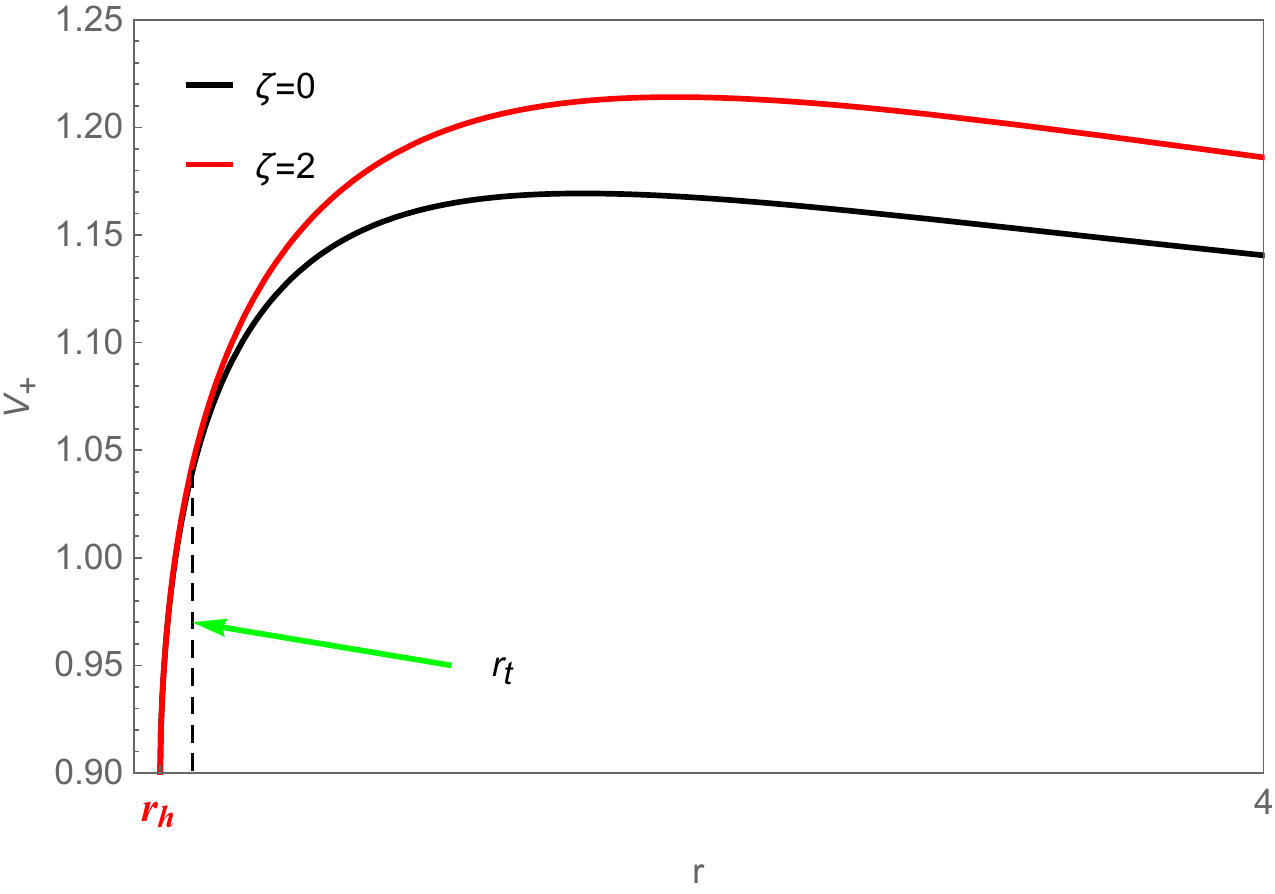}}
	\subfigure[particle 2]{\includegraphics[width=0.329\textwidth,height=4.5cm,keepaspectratio]{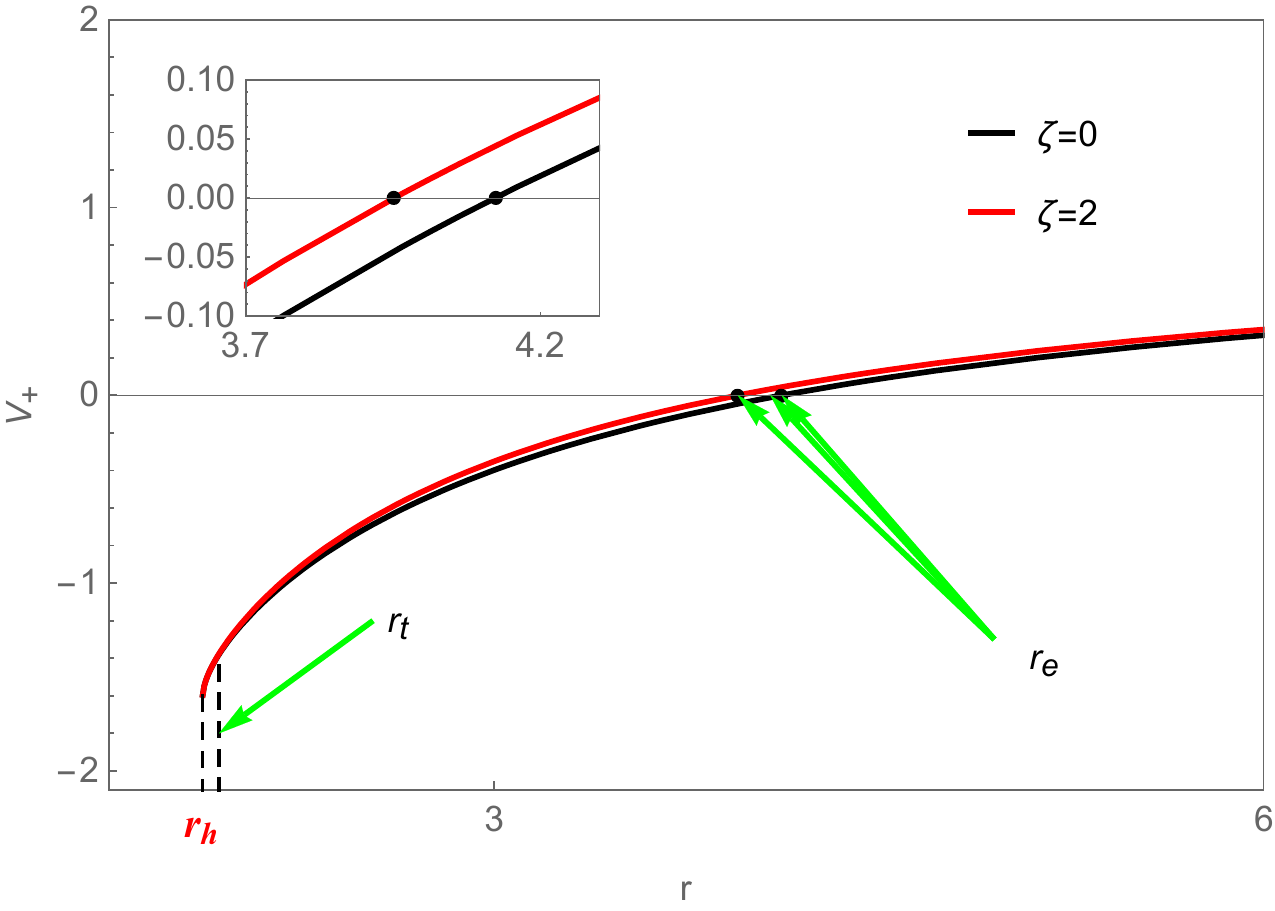}}
	\subfigure[particle 3]{\includegraphics[width=0.329\textwidth,height=4.5cm,keepaspectratio]{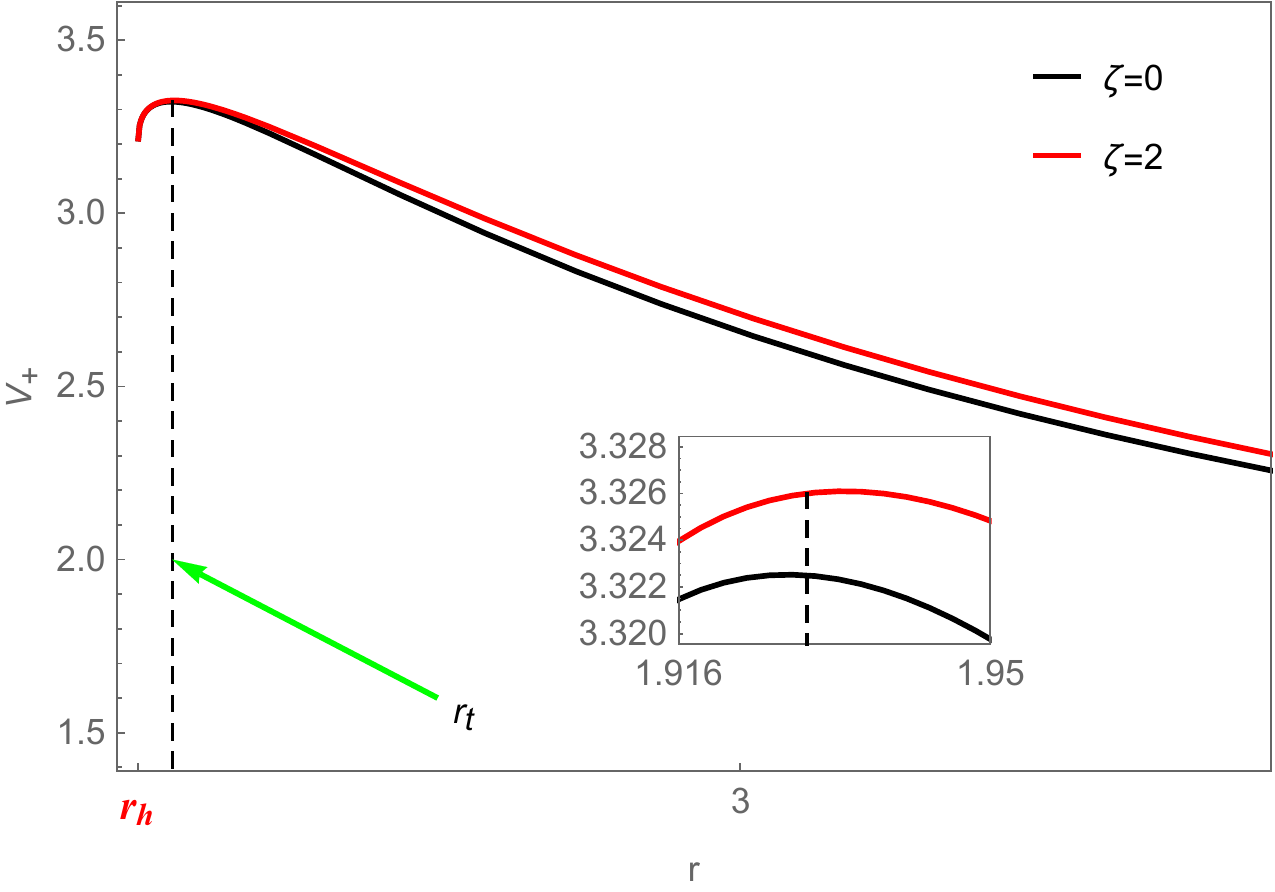}}

	\caption{Effective potentials $V_+$ of particles 1, 2, and 3 for different values of the quantum parameter $\zeta$, at the turning point $r_t=1.93$.}
	\label{fig_half}
\end{figure*}

\begin{figure*}[htbp]
	\centering
	\subfigure[RN BH]{\includegraphics[width=0.49\textwidth, height=6in, keepaspectratio]{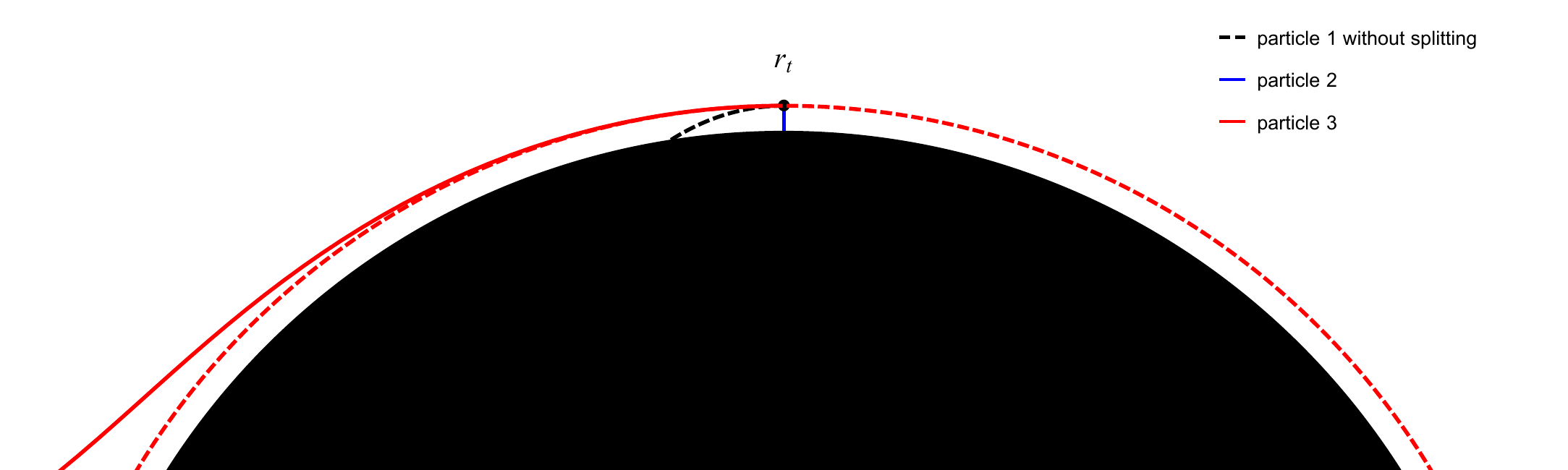}}
	\subfigure[quantum-corrected BH with $\zeta=2$]{\includegraphics[width=0.49\textwidth,height=6in, keepaspectratio]{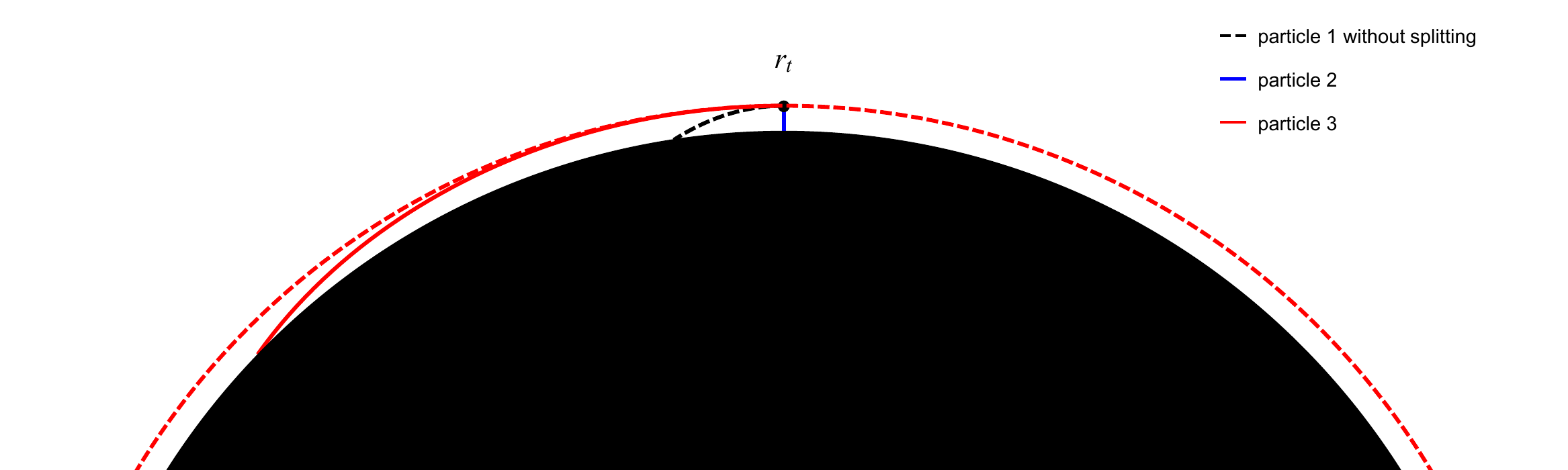}}

	\caption{Partial particle trajectories at $r_t = 1.93$ for the RN BH and the quantum-corrected BH with $\zeta=2$. The red dashed line represents a circle of radius $r_t$, and the black dashed curve shows the trajectory of particle 1 without splitting. The blue and red solid curves correspond to the trajectories of particles 2 and 3, respectively. The black region indicates a partial cross‑section of the BH.}
	\label{fig_half_guiji}
\end{figure*}

As an example using the initial parameters from table~\ref{tab:1} and $Q = 0.5$, figure~\ref{fig_half} shows the effective potentials at $r_t = 1.93$ for $\zeta = 0$ and $\zeta = 2$. Here, for $\zeta = 0$, the turning point $r_t$ lies between the peak coordinate $r_m^{(3)}$ of the RN case and the peak coordinate $R_m^{(3)}$ corresponding to $\zeta = 2$ (i.e., $r_m^{(3)} < r_t < R_m^{(3)}$). The trajectories of particles 2 and 3 resulting from the splitting of particle 1 at this turning point are displayed in figure~\ref{fig_half_guiji}. The results are consistent with the discussion above, showing that under specific conditions, the trajectory of particle 3 near a quantum-corrected BH can differ from that near a classical RN one: while it may escape the RN BH, it can become trapped in the quantum-corrected case. To further investigate whether the qualitative influence of the quantum parameter persists over a broader parameter range, under the same mass and charge parameter values, we present the post‑extraction particle trajectories for the case $L_2 \neq 0$ (taking $L_2 = 1$ and $\sigma = 1$ as an example). The results confirm that the qualitative effect on particle 3 remains valid. The changes in particle escape behavior induced by the introduction of quantum corrections may affect the upper limit, spectral energy distribution, or luminosity of high-energy radiation near the BH. In the future, combining light-curve analysis with spectral fitting could help distinguish between the classical RN BH and its quantum-corrected counterpart.

\begin{figure*}[htbp]
	\centering
	\subfigure[RN BH]{\includegraphics[width=0.49\textwidth, height=6in, keepaspectratio]{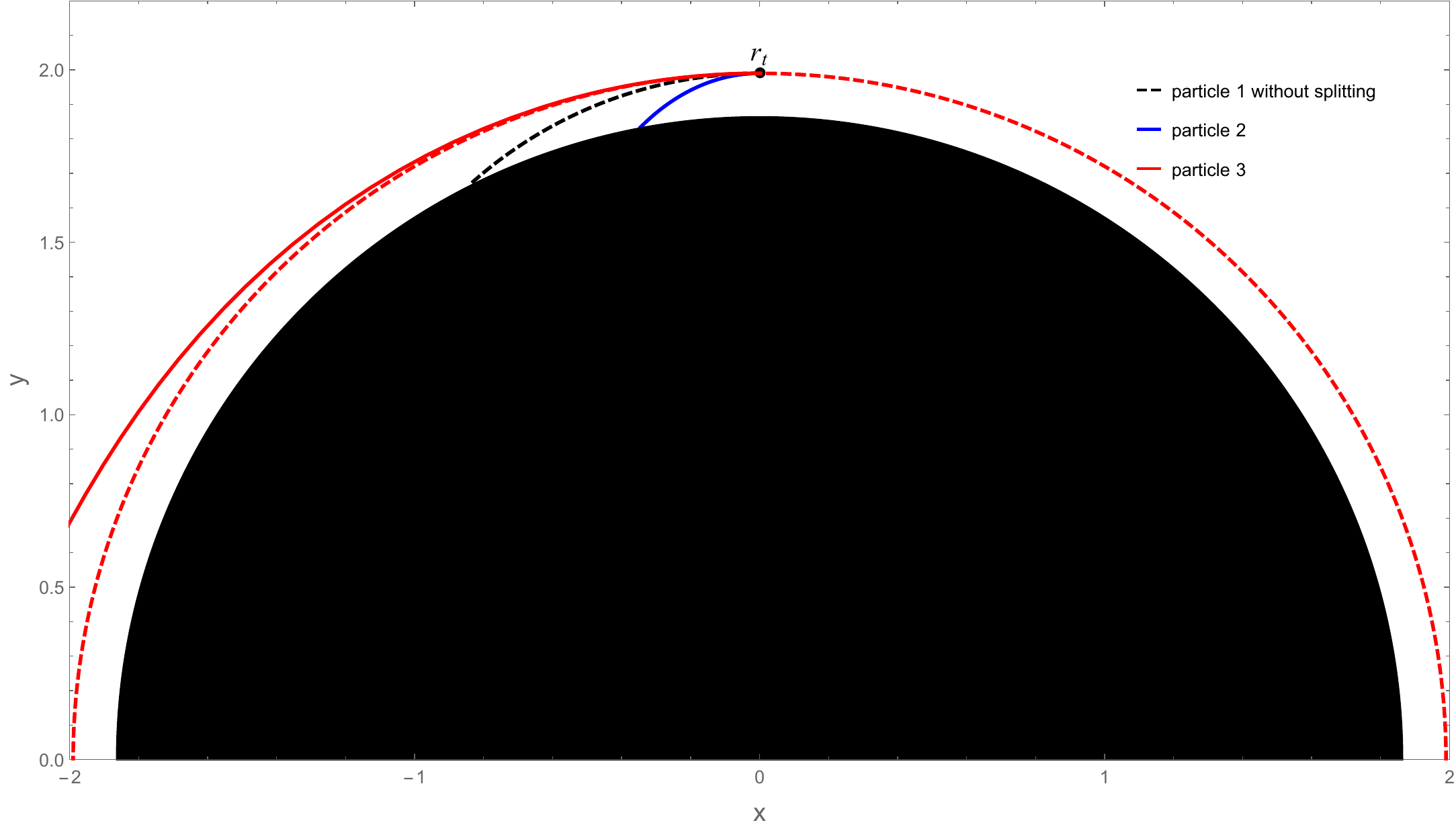}}
	\subfigure[quantum-corrected BH with $\zeta=2$]{\includegraphics[width=0.49\textwidth,height=6in, keepaspectratio]{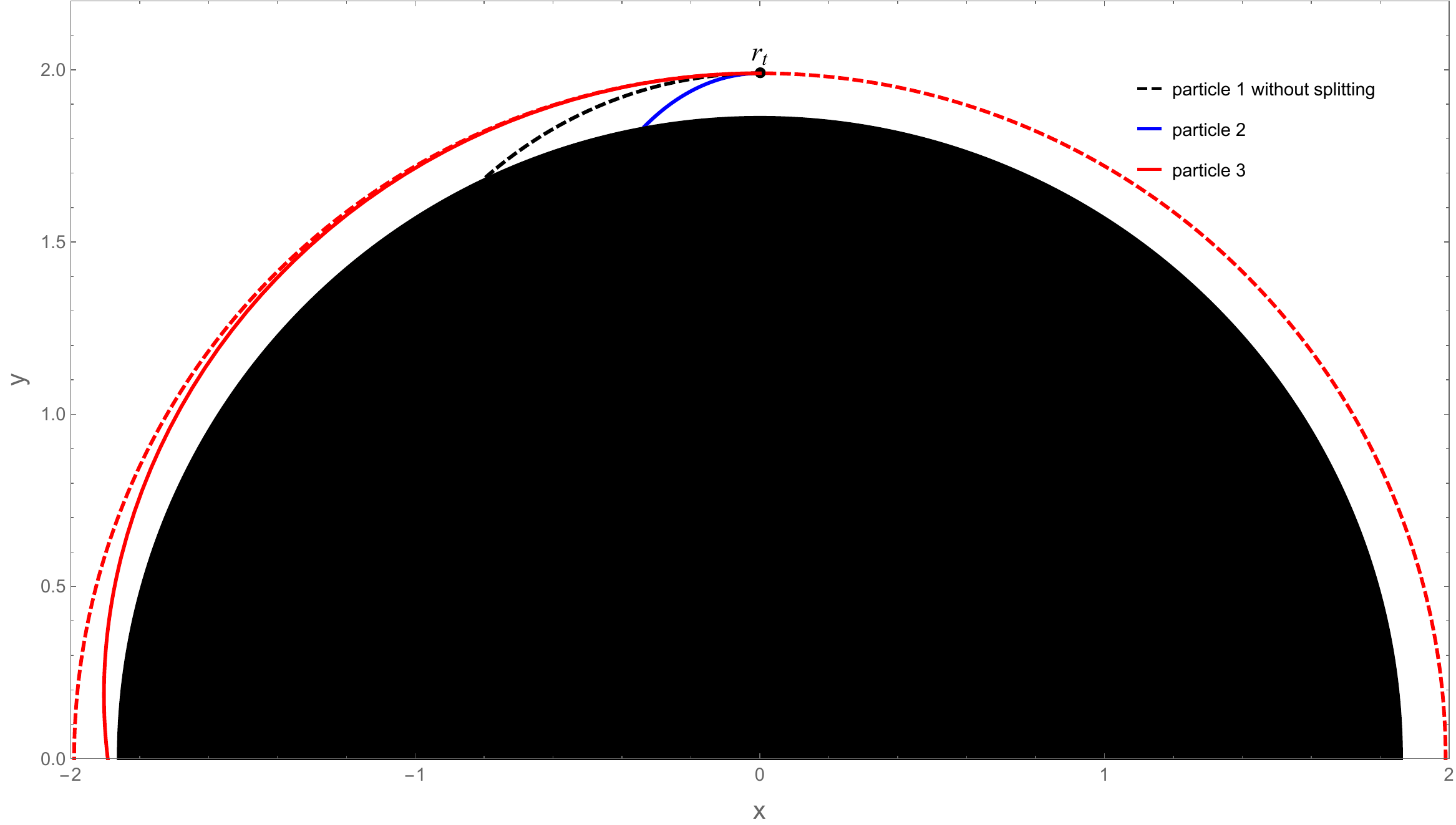}}

	\caption{Partial particle trajectories at the turning point $r_t = 1.99$ with $L_2 = 1$ for the RN BH and the quantum‑corrected BH ($\zeta = 2$). The red dashed line denotes the circle of radius $r_t$, and the black dashed curve shows the trajectory of particle 1 without splitting. The blue and red solid curves correspond to the trajectories of particles 2 and 3, respectively. The black region indicates a partial cross‑section of the BH.}
	\label{fig_half_guiji2}
\end{figure*}

\section{Summary}\label{section6}

In this paper, we studied the electric Penrose process for charged particles around a covariant quantum-corrected RN BH. We first provided a brief review of this quantum-corrected BH and discussed its horizons, as well as the constraints on the BH parameters. Subsequently, we analyzed the motion of charged particles around this BH, derived the equations of motion and the effective potential, and examined how the relative magnitude between the effective potential and the particle energy determines the particle's motion. We also explored the influence of the quantum parameter $\zeta$ on the effective potential. It turns out that increasing $\zeta$ enhances the peak value of the effective potential.

Within the spacetime background of the quantum-corrected BH, we then derived the generalized ergoregion condition for charged particle energy extraction and systematically analyzed how particle charge, angular momentum, and the quantum parameter $\zeta$ affect the ergoregion boundary $r_e$. Our findings reveal that, for fixed BH parameters, the generalized ergoregion boundary expands outward with increasing magnitude of the particle's negative charge, and grows significantly as the absolute value of its angular momentum decreases. Meanwhile, the ergoregion boundary gradually contracts as $\zeta$ increases. This indicates that, under identical conditions, the effective region for energy extraction in the classical RN BH is larger than that in its quantum-corrected counterpart.

Subsequently, in a further exploration of the Penrose energy extraction mechanism, we analyzed the simplified scenario where the splitting point coincides with the turning points of all particles. By establishing the conservation equations among the particles, we derived a general expression for the energy extraction efficiency [eq.~\eqref{xiaolv}]. Under fixed initial parameters, we analyzed the variation of the energy extraction efficiency $\eta$ with the BH charge $Q$ and the quantum parameter $\zeta$. The results show that $\eta$ increases with $Q$ and decreases monotonically with $\zeta$; when $\zeta$ exceeds a certain critical value, $\eta$ even approaches zero. This phenomenon originates from the gradual inward contraction of the ergoregion boundary $r_e$ caused by increasing $\zeta$, until $r_e$ becomes smaller than the particle turning point $r_t$. At this point, the generalized ergoregion condition cannot be satisfied, rendering an effective energy extraction process impossible.

In addition, we examined the subsequent motion of particle 3 resulting from the splitting of particle 1 (originating from a distant observer) in a general electric Penrose process. For the three cases, (i) $L_2 = 0$, (ii) $L_2 > 0$ with $\sigma = 1$, and (iii) $L_2 < 0$ with $\sigma = -1$, we rigorously proved that the resulting particle 3 will always carry greater energy back to a distant observer. Although we have not obtained a general proof that particle 3 can escape in all cases, our proof here represents an extension of previous discussions. The conclusions obtained under this framework are applicable to particle motion in a wide range of charged BH models. As an illustrative example, the trajectory results in figure~\ref{fig_guiji} correspond to the parameters in table~\ref{tab:1}. The results demonstrate that particle 3 moves away from the BH for all values of the quantum parameter $\zeta$, with $\zeta$ exerting only a weak modulation on its trajectory shape.

Finally, we systematically investigated particle motion in a special electric Penrose process, where the turning point of particle 1 lies inside the effective-potential peak coordinate $r_m$ and it cannot escape the BH. We find that even though particle 1 itself is unable to escape the BH, the fragment particle 3 produced by its splitting at the turning point can exhibit two distinct dynamical outcomes: successful escape from the BH (as shown in figure~\ref{fig_guiji1}), or remain trapped (as depicted in figure~\ref{fig_no}). The difference between these behaviors is governed primarily by the distance of the splitting point $r_t$ relative to the horizon, while the influence of the quantum parameter $\zeta$ remains generally weak. However, we uncover a key phenomenon: under certain critical conditions, the presence of $\zeta$ can qualitatively alter the fate of particle 3. Specifically, particle 3, which escapes in the classical RN BH, may become unable to escape in its quantum-corrected counterpart (figure~\ref{fig_half_guiji}), and this also occurs for $L_2>0$ (figure~\ref{fig_half_guiji2}). This behavioral transition stems from a subtle shift in the peak position $r_m$ of the effective potential of particle 3 caused by increasing $\zeta$, which alters the accessible region for particle motion around the BH and ultimately changes its outcome. The observed differences in charged-particle trajectories during the electric Penrose process offer a potential kinematic signature for distinguishing the classical RN BH from its quantum-corrected counterpart.

In conclusion, this study provides a systematic analysis of the electric Penrose process for charged particles in the spacetime of a quantum-corrected RN BH, revealing the suppressive effect of the quantum parameter $\zeta$ on the process. Specifically, the introduction of $\zeta$ leads to the contraction of the generalized ergoregion boundary $r_e$, a decrease in the energy extraction efficiency $\eta$, and --- under specific parameter conditions --- a transition of the energy-gaining particle 3 from an escapable to a non-escapable state. Although the analysis in this paper is based on test-particle kinematics in a fixed background, the obtained results can still provide qualitative insights for high-energy astrophysical observations. Recently, Zhao et al. systematically demonstrated, for the first time, the imaging and light-curve characteristics of hot-spot splitting in the magnetic-reconnection-driven Penrose process~\cite{Zhao:2025ouq}. They proposed a method to distinguish the infall of negative-energy and positive-energy fragments via flare patterns, thereby providing specific theoretical foundations and observable signatures for the observational identification of the Penrose process. Inspired by this line of thought, the reduced energy extraction efficiency of the Penrose process in the quantum‑corrected BH studied here (figure~\ref{fig_eta}), together with the fate transition of high‑energy fragments from escape to capture under specific conditions (figure~\ref{fig_half_guiji}), is expected to directly affect the Doppler blueshift, the trajectory, and the temporal evolution. Consequently, these effects will leave identifiable imprints on the light curves (flare number and intensity) as well as on the centroid trajectory. Therefore, analyzing flare events near BHs with next‑generation high‑resolution observational facilities holds promise for providing potential observational evidence to distinguish between the classical RN BH and its quantum‑corrected counterpart.

It must be acknowledged that, for the sake of theoretical tractability and clarity, the present analysis has certain limitations. On the one hand, we have not considered quantum‑corrected BH spacetimes with a cosmological constant $\Lambda$; a straightforward extension to include $\Lambda$ is natural and feasible. On the other hand, the splitting point is assumed to coincide with the common turning point of all three particles, and the proof of particle escape as well as the subsequent motion analysis are based on this simplified condition. Future work will be devoted to proving particle escape under more general conditions, further analyzing the energy extraction process when the splitting point does not coincide with the turning point, and systematically investigating the light‑curve modulations and flare features associated with the energy extraction process. This will provide richer observational criteria for distinguishing quantum‑corrected BHs from their classical counterparts.

%-------------------------------------------------------------
\acknowledgments This work is supported in part by NSFC Grants No. 12165005 and No. 11961131013.
%-------------------------------------------------------------

\appendix

\section{Proof of the escape of particle 3 from a BH}\label{appendix}

In section~\ref{section4}, we have already established that $V_{3}' < 0$ when $G_{1}(r) \leq 0$. In what follows we derive the sign of $V_{3}'$ for the case $G_{1}(r) > 0$. Hereafter, we use the shorthand $G_1 \equiv G_{1}(r)=r_t f'(r_t) - 2 f(r_t)$.

For notational clarity, we denote the terms in eqs.~\eqref{Eq:V1} and \eqref{Eq:V3} that determine the sign of the effective potential as $T_1$ (for particle 1) and $T_2$ (for particle 3), i.e.,
\begin{align}
	T_1&\equiv-2q_1 Q S_1+L_1^2 G_1+H,	\label{Eq_A10} \\
	T_2&\equiv-2q_3 Q S_2+ M_1^2 L_1^2 G_1+H,
	\label{Eq_A1}
\end{align}
where
\begin{align}
	S_1&=\sqrt{\left(L_1^2+r_t^2\right) f(r_t)},\\
	S_2&=\sqrt{\left(M_1 ^2 L_1^2 + r_t^2\right) f(r_t)},\\
	H&=r_t^3 f'(r_t).
	\label{Eq_A2}
\end{align}
Combining the expressions for $H$ and $G_1$, we obtain
\begin{align}
	H&=r_t^2\left(G_1+2f(r_t)\right).
	\label{Eq_A3}
\end{align}

From $T_1 < 0$, we can deduce that $H$ must satisfy
\begin{align}
	0<H<2q_1 Q S_1-L_1^2 G_1.
	\label{Eq_A4}
\end{align}
Putting eq.~\eqref{Eq_A4} into eq.~\eqref{Eq_A1}, we obtain
\begin{align}
	T_2<-2Q (q_3 S_2 -q_1 S_1)+ (M_1^2 -1) L_1^2 G_1.
	\label{Eq_A5}
\end{align}
A negative right-hand side in eq.~\eqref{Eq_A5} implies $T_2 < 0$. Therefore, the condition $T_2 < 0$ is guaranteed if the following inequality holds:
\begin{align}
	(M_1^2 - 1) L_1^2 G_1 < 2 q_1 Q S_1 (R K - 1),
	\label{Eq_A6}
\end{align}
where we have introduced the shorthand notations
\begin{align}
	R \equiv \frac{q_3}{q_1}\,, \qquad
	K \equiv \sqrt{\frac{(M_1^2 L_1^2 + r_t^2) f(r_t)}{(L_1^2 + r_t^2) f(r_t)}}.
	\label{Eq_A7}
\end{align}
Under the condition $T_1 < 0$, an upper bound for $G_1$ can also be derived using eqs.~\eqref{Eq_A10} and \eqref{Eq_A3}, namely,
\begin{equation}
	0 < G_1 < \frac{2 q_1 Q S_1 - 2 r_t^2 f(r_t)}{r_t^2 + L_1^2}.
	\label{eq_bound}
\end{equation}
Since $2 r_t^2 f(r_t) > 0$ outside the event horizon, we obtain:
\begin{equation}
	0 < G_1 < \frac{2 q_1 Q S_1}{r_t^2 + L_1^2}.
	\label{eq_bound1}
\end{equation}
If this upper bound satisfies
\begin{equation}
	L_1^2(M_1^2 - 1) \frac{2 q_1 Q S_1}{r_t^2 + L_1^2} < 2 q_1 Q S_1 (R K - 1),
	\label{eq_bound2}
\end{equation}
then eq.~\eqref{Eq_A6} is automatically satisfied. In general, $2 q_1 Q S_1 > 0$. After simplification, we obtain
\begin{equation}
	\frac{L_1^2(M_1^2 - 1)}{r_t^2 + L_1^2} < R K - 1.
	\label{eq_bound3}
\end{equation}
Together with eq.~\eqref{eq_qm}, this leads directly to
\begin{equation}
	M_1 K - 1 < R K - 1.
	\label{eq_RK}
\end{equation}
Likewise, if the inequality
\begin{equation}
	\frac{L_1^2(M_1^2 - 1)}{r_t^2 + L_1^2} < M_1 K - 1
	\label{eq_boun4}
\end{equation}
holds, then eq.~\eqref{eq_bound3} is automatically satisfied. Inserting the explicit form of $K$ from eq.~\eqref{Eq_A7} and setting
\begin{equation}
	U \equiv \frac{L_1^2(M_1^2 - 1)}{r_t^2 + L_1^2},
	\label{eq_boun44}
\end{equation}
we can rewrite eq.~\eqref{eq_boun4} as
\begin{equation}
	U < M_1 \sqrt{U+1} - 1.
	\label{eq_boun5}
\end{equation}
Since $U \geq 0$ is evident, further simplification yields
\begin{equation}
	U+1 < M_1 ^2.
	\label{eq_boun6}
\end{equation}
When $L_1 = 0$, eq.~\eqref{eq_boun44} gives $U = 0$, while $M_1 > 1$ is self-evident. Hence eq.~\eqref{eq_boun6} is satisfied. For $L_1 \neq 0$, eq.~\eqref{eq_boun44} reads
\begin{equation}
	U r_t ^2+(U+1)L_1^2 = M_1 ^2 L_1^2.
	\label{eq_boun7}
\end{equation}
It is straightforward to verify that eq.~\eqref{eq_boun6} holds universally. Thus, we have demonstrated that $T_2 < 0$ remains true even for $G_{1}(r) > 0$. This completes the proof that particle 3, produced from the split of particle 1 at a turning point near the BH after its journey from a distant observer, can return to the observer with a net gain in energy.

\section{Partial proof for particle 3 escape when $L_2 \neq 0$}\label{appendix1}

We further consider a more general case, namely, when particle 1 splits into particles 2 and 3 at the turning point, with the angular momentum of particle 2 satisfying $L_2 \neq 0$ (here we mainly discuss the case $L_2>0$; the situation for $L_2<0$ is similar). For $L_2>0$, it follows from eqs.~\eqref{L0} and \eqref{L3} that there are two possible values for $L_1$ and $L_3$, depending on the value of $\sigma$ ($\sigma = 1$ or $\sigma = -1$).

We first consider the case $\sigma = 1$. Combining eqs.~\eqref{L0} and \eqref{L3}, we have
\begin{align}
	L_3-L_1 =\frac{1}{2m_2}\left[L_2\frac{m_1 C_2-m_3 C_1}{m_1 m_3}+\sqrt{C_3 \left(L_2^2 + r_t^2\right)}\left(\frac{1}{m_3}-\frac{1}{m_1}\right)\right].
	\label{Eq_B0}
\end{align}
Substituting eqs.~\eqref{c1} and \eqref{c2} into $m_1 C_2 - m_3 C_1$, we have
\begin{align}
	m_1 C_2 - m_3 C_1
	&= m_1 (m_1^2 - m_2^2 - m_3^2) - m_3 (m_1^2 + m_2^2 - m_3^2) \nonumber\\
	&= m_1^3 - m_1 m_2^2 - m_1 m_3^2 - m_3 m_1^2 - m_3 m_2^2 + m_3^3 \\
	&= (m_1 + m_3)\bigl[(m_1 - m_3)^2 - m_2^2\bigr] \nonumber.
\end{align}
Since $m_1 \geq m_2 + m_3$, the above expression is necessarily greater than or equal to zero. Therefore, in this case, eq.~\eqref{Eq_B0} is certainly nonnegative, and we have $L_3 \geq L_1>0$.

Subsequently, we rewrite eq.~\eqref{Eq_A1} as
\begin{align}
	T_3&\equiv-2q_3 Q S_3+ L_3^2 G_1+H,
	\label{Eq_B1}
\end{align}
where
\begin{align}
	S_3&=\sqrt{\left( L_3^2 + r_t^2\right) f(r_t)}.
	\label{Eq_B2}
\end{align}
For the case $L_3 = L_1$, it is evident that $T_3 < 0$, and particle 3 can naturally escape the BH. For the case $L_3 > L_1$, if $G_1 < 0$, then it is also clear that $T_3 < 0$. Therefore, we only need to discuss the sign of $T_3$ for the case $L_3 > L_1$ and $G_1 > 0$. Given that $T_1 < 0$, if in this case $T_3 - T_1 < 0$, this implies $T_3 < 0$, and thus we have
\begin{align}
	\Delta \equiv T_3-T_1=-2Q (q_3 S_3 -q_1 S_1)+ (L_3^2-L_1^2) G_1<0.
	\label{Eq_B3}
\end{align}
For eq.~\eqref{Eq_B3} to hold, the following condition must be satisfied:
\begin{align}
	(L_3^2-L_1^2) G_1<2Q q_1 S_1 (R \frac{S_3}{S_1} -1).
	\label{Eq_B4}
\end{align}
Here, $R$ is a shorthand notation we have used from eq.~\eqref{Eq_A7}. Using the upper bound of $G_1$ from eq.~\eqref{eq_bound1}, the above expression simplifies to
\begin{align}
	(L_3^2-L_1^2) \frac{2 q_1 Q S_1}{r_t^2 + L_1^2}<2Q q_1 S_1 (R \frac{S_3}{S_1} -1).
	\label{Eq_B5}
\end{align}
If eq.~\eqref{Eq_B5} holds, then eq.~\eqref{Eq_B4} naturally holds as well. Then, we have
\begin{align}
	\frac{L_3^2-L_1^2}{r_t^2 + L_1^2}< R \sqrt{\frac{( L_3^2 + r_t^2) f(r_t)}{(L_1^2 + r_t^2) f(r_t)}} -1.
	\label{Eq_B6}
\end{align}
After simplification, we obtain
\begin{align}
	\frac{L_3^2-L_1^2}{r_t^2 + L_1^2}< R \sqrt{\frac{L_3^2-L_1^2}{r_t^2 + L_1^2}+1} -1.
	\label{Eq_B7}
\end{align}
We denote $X \equiv \frac{L_3^2-L_1^2}{r_t^2 + L_1^2}$. Since $X>0$, further simplification yields
\begin{align}
	X< R^2-1.
	\label{Eq_B8}
\end{align}
As previously known, $R > M_1$; therefore, when $X < M_1^2 - 1$, the above relation naturally holds. Combining eq.~\eqref{L_momentm} with the condition $L_3 > L_1>0$, we find that $L_3 / L_1 < M_1$. Then $X$ satisfies
\begin{align}
	X = \frac{L_3^2-L_1^2}{r_t^2 + L_1^2}=\frac{\frac{L_3^2}{L_1^2}-1}{\frac{r_t^2}{L_1^2} + 1} < \frac{M_1^2-1}{\frac{r_t^2}{L_1^2} + 1}<M_1^2-1<R^2-1.
	\label{Eq_B9}
\end{align}
Thus, we have proved that under the conditions $L_3 > L_1$ and $L_2 > 0$, particle 3, which is produced from the splitting of particle 1, can always gain more energy and return to the observer.

For the case $\sigma = -1$, when $L_2$ satisfies $L_2 C_1 < \sqrt{C_3 \bigl(\mathcal{E}_2^2 - f(r_t)\bigr)}$ (in which case both $L_3$ and $L_1$ are negative), one can repeat the above steps to show that $|L_3| > |L_1| > 0$. However, since $|L_3|/|L_1| > M_1$ in this case, a proof that particle 3 always escapes cannot be obtained. For the opposite case where $L_2 C_2 > \sqrt{C_3 \bigl(\mathcal{E}_2^2 - f(r_t)\bigr)}$, both $L_3$ and $L_1$ are positive, and $L_3 - L_1$ can be written as
\begin{align}
	L_3-L_1 =\frac{1}{2m_1 m_2 m_3}\left[L_2( m_1 C_2-m_3 C_1)-\sqrt{C_3 \left(L_2^2 + r_t^2\right)}\left(m_1-m_3\right)\right].
	\label{Eq_s1}
\end{align}
For the term $\sqrt{C_3 \left(L_2^2 + r_t^2\right)}\left(m_1-m_3\right)$, it satisfies
\begin{align}
	\sqrt{C_3 \left(L_2^2 + r_t^2\right)}\left(m_1-m_3\right)>L_2 \sqrt{C_3}\left(m_1-m_3\right)>0.
	\label{Eq_s2}
\end{align}
Since we have
\begin{align}
	C_3 \left(m_1-m_3\right)^2 - \left(m_1 C_2 - m_3 C_1\right)^2 = 4 m_1 m_2^2 m_3 \big[(m_1 - m_3)^2 - m_2^2\big] > 0,
	\label{Eq_s3}
\end{align}
it follows that $\sqrt{C_3 \left(L_2^2 + r_t^2\right)}\left(m_1-m_3\right) > L_2( m_1 C_2 - m_3 C_1)$. This indicates that for $\sigma = -1$, the case $L_3 < L_1$ occurs. For $L_3 < L_1$, the sign of $T_3$ can also be determined by the sign of $\Delta$. The conditions for $\Delta < 0$ are as follows.
\begin{itemize}
	\item When $q_3 S_3 - q_1 S_1 > 0$: either $G_1 > 0$, or $G_1 < 0$ with $2Q(q_3 S_3 - q_1 S_1) > (L_3^2 - L_1^2)G_1$;
	\item When $q_3 S_3 - q_1 S_1 = 0$: $G_1 > 0$;
	\item When $q_3 S_3 - q_1 S_1 < 0$: $G_1 > 0$ and $2Q(q_3 S_3 - q_1 S_1) < (L_3^2 - L_1^2)G_1$.
\end{itemize}
It remains unproven that the above conditions hold for arbitrary initial parameters of the particles. This means that, in the case $L_3 < L_1$, whether particle 3 can necessarily gain more energy and escape the BH remains unknown; it can only be determined by specifying the concrete initial conditions.

Next, we briefly investigate the escape of particles under the condition $L_2 < 0$. From eqs.~\eqref{L0} and \eqref{L3}, it follows that when $\sigma = -1$, both $L_1$ and $L_3$ are negative. Similar to the above derivation, it can be proved that
\begin{align}
	|L_3|-|L_1|>0.
	\label{Eq_l2}
\end{align}
Likewise, we can obtain $L_3/L_1 < M_1$ from eq.~\eqref{L_momentm}. Hence, the above derivation still applies here: for $L_2 < 0$ and $\sigma = -1$, particle 3 is guaranteed to satisfy the escape condition $T_3<0$. However, for the case $\sigma = 1$ (which is similar to the case $L_2 > 0$ and $\sigma = -1$), we are still unable to obtain a general proof.

Up to now, we can only conclude that under the conditions $L_2 > 0$ and $\sigma = 1$, or $L_2 < 0$ and $\sigma = -1$, particle 3 is guaranteed to escape the BH. However, under other conditions ($ L_2 >0 $ and $\sigma=-1$, or $L_2 < 0$ and $\sigma = 1$), a general proof of the escape of particle 3 remains unattainable and requires further analysis based on the specific initial conditions of the BH and particle system. This issue can only be left for future research.

%%=============================================================================================

%%=============================================================================================

\end{document}